\documentclass[useAMS,usenatbib]{mn2e}

\usepackage[latin1]{inputenc}  
\usepackage{graphicx}
\usepackage{txfonts}

\title[Cosmic ray induced interstellar unsaturated hydrocarbons]{Formation of unsaturated hydrocarbons in interstellar ice analogs \\ by
cosmic rays}

\author[S. Pilling et al.]{S. Pilling$^{1}$\thanks{E-mail:
sergiopilling@pq.cnpq.br}, D. P. P. Andrade$^{1}$, E. F. da Silveira$^{2}$, \and H. Rothard$^{3}$, A. Domaracka$^{3}$, P. Boduch $^{3}$\\
\\
$^{1}$Universidade do Vale do Paraíba (UNIVAP), Instituto de Pesquisa e Desenvolvimento (IP\&D), São José dos Campos, SP, Brazil.\\
$^{2}$Pontifícia Universidade Católica do Rio de Janeiro (PUC-Rio), Instituto de Fisica, Rio de Janeiro, RJ, Brazil.\\
$^{3}$Centre de Recherche sur les Ions, les Matériaux et la Photonique CIMAP (GANIL/CEA/CNRS/ENSICAEN/Université de Caen Basse-Normandie), Caen, France.\\}

\begin{document}

\date{Received / Accepted}

\pagerange{\pageref{firstpage}--\pageref{lastpage}} \pubyear{2005}

\maketitle

\label{firstpage}


\begin{abstract} 

The formation of C=C and C$\equiv$C bonds from the processing of pure c-C$_6$H$_{12}$ (cyclohexane) and mixed H$_2$O:NH$_3$:c-C$_6$H$_{12}$ (1:0.3:0.7) ices by highly-charged, and energetic ions (219 MeV $^{16}$O$^{7+}$ and 632 MeV $^{58}$Ni$^{24+}$) is studied. The experiments simulate the physical chemistry induced by medium-mass and heavy-ion cosmic rays in interstellar ices analogs. The measurements were performed inside a high vacuum chamber at the heavy-ion accelerator GANIL (Grand Accelératéur National d'Ions Lourds) in Caen, France. The gas samples were deposited onto a polished CsI substrate previously cooled to 13 K. In-situ analysis was performed by a Fourier transform infrared (FTIR) spectrometry at different ion fluences. Dissociation cross section of cyclohexane and its half-life in astrophysical environments were determined.
A comparison between spectra of bombarded ices and young stellar sources indicates that the initial composition of grains in theses environments should contain a mixture of H$_2$O, NH$_3$, CO (or CO$_2$), simple alkanes, and CH$_3$OH. Several species containing double or triple bounds were identified in the radiochemical products, such as hexene, cyclohexene, benzene, OCN$^-$, CO, CO$_2$, as well as several aliphatic and aromatic alkenes and alkynes. The results suggest an alternative scenario for the production of unsaturated hydrocarbons and possibly aromatic rings (via dehydrogenation processes) in interstellar ices induced by cosmic ray bombardment.
\end{abstract}


\section{Introduction}
A large number of gas-phase molecules containing up to 13 atoms has been discovered in interstellar space (see review in Herbst \& van Dishoeck 2009; Ehrenfreund \& Charnley 2000). These interstellar molecules are mainly organic compounds involving unsaturated carbon-chain species such as simple alkenes (C$_n$H$_{2n}$), cumulene carbenes (H$_2$C(=C)$_n$), and alkynes (C$_n$H$_{2n-2}$) (e.g. Thaddeus \&  McCarthy 2011; Herbst 1995; Winnewisser \& Herbst 1987). Cumulene carbenes and other unsaturated carbon-chain species have also been suggested in literature as possible carriers for diffuse interstellar bands (DIBs) (Apponi et al. 2000).

Unsaturated small compounds such C$_ 4$H$_2$, C$_6$H$_2$, methylpolyynes (e.g. CH$_3$C$_2$H and CH$_3$C$_4$H) and also ring-compounds such as benzene have also been detected in space environments such as in dust shells of late stars, proto-planetary nebulaes (Cernicharo et al. 2001a, 2001b). Long aliphatic unsaturated chains containing nitrogen atoms, such as polyacetylenes or polyynes H-(C$\equiv$C)$_n$-H, and cyanopolyynes (H-(C$\equiv$C)$_n$-CN), may also be present in interstellar medium (Fukuzawa et al. 1998). Polyacetylenes are considered the missing link between small gas phase molecules, among them acetylene (C$_2$H$_2$), carbon monoxide (CO), and hydrogen cyanide (HCN), and carbonaceous grain particles formed in the outflow of carbon-rich stars (Cordiner \& Millar  2009; Duley et al. 2005). Aromatic compounds such as naphthalene and anthracene cation have been detected in cold molecular clouds (Iglesias-Groth et al. 2008, 2010). In addition, Duley et al. (2005) suggested the presence of cyclic alkanes and alkenes, such as cyclohexene, in attempt to justify the appearance of spectral features near 3.4 $\mu$m in the infrared spectra of interstellar clouds (IC) and in diffuse interstellar medium (DISM). For example, they found that the average CH$_2$/CH$_3$ ratio in a number of cyclohexene and cyclohexadiene compounds with spectra that match the 3.4 $\mu$m band in the proto-planetary nebula (PPN) CRL 618 is $\approx$1.15. Single ring systems, such as c-C$_3$H and c-C$_3$H$_2$, have also been detected in space (Smith 1992).

Several mechanisms have been proposed to explain the presence of these unsaturated compounds in space including gas-phase ion-molecule reactions and neutral-neutral reactions (e.g. Cherchneff et al. 1992; Taylor \& Duley 1997; Fukuzawa et al. 1998; Duley et al. 2005, and references therein). The present experimental radiochemical study suggests an alternative scenario for the production of unsaturated carbon chain species (and dehydrogenation) in interstellar ices induced by cosmic rays bombardment. The experiments were performed bombarding cyclohexane (c-C$_6$H$_{12}$)-containing ices at 13 K with fast ions to simulate the physical chemistry induced by cosmic rays in interstellar unsaturated ice analogs inside dense regions of interstellar medium, such as molecular clouds and protostellar disks. Although no large saturated cyclic compound in the interstellar medium, such as cyclohexane, has been direct detected, the presence of this compound, as well as of other large saturated hydrocarbons has been suggested in these regions (e.g. Pendleton et al. 1994; Duley et al. 2005). In this paper, this compound is just a prototype for saturated hydrocarbons. We expect that the unsaturation processes described here also occur with other saturated hydrocarbons already detected in space(e.g. Pendleton \& Allamandola 2002; Duley et al. 2006). In addition, saturated hydrocarbons can be formed on the surface of frozen grain by H-atom reaction on graphite grains (Bar-Num et al. 1980) and direct hydrogenation (catalytic hydrogenation) of alkenes/alkynes (e.g. Olah \& Molnár 2003; Marcelino et al. 2007).

Section 2 describes briefly the experimental setup. The results on the radiolysis of saturated hydrocarbon containing ices, including formation and dissociation cross sections, are presented and discussed in section 3. Some astrophysical implications, as well as an estimate for half-lives of these molecules at possible astrophysical environments, are provided and discussed in section 4. Finally, section 5 contains the final remarks and conclusions.

\section{Experimental}

To simulate the physico-chemical changes induced by medium-mass and heavy-ion cosmic rays in interstellar ices analogs, we used the facilities of the heavy-ion accelerator GANIL (Grand Accelératéur National d'Ions Lourds) in Caen, France. The measurements were performed inside a high vacuum chamber which could be coupled to different ion beamlines: IRRSUD (for projectiles with 219 MeV $^{16}$O$^{7+}$) and SME (for 632 MeV $^{58}$Ni$^{24+}$). The beam flux was $\phi =2 \times 10^{9}$ ions cm$^{-2}$ s$^{-1}$. The gas samples were deposited onto a polished CsI substrate previously cooled to 13 K. The ion projectiles impinged perpendicularly onto the ice target. In-situ analysis was performed by a Fourier transform infrared (FTIR) spectrometry at different ion fluences.

\begin{figure}
 \centering
 \includegraphics[scale=0.75]{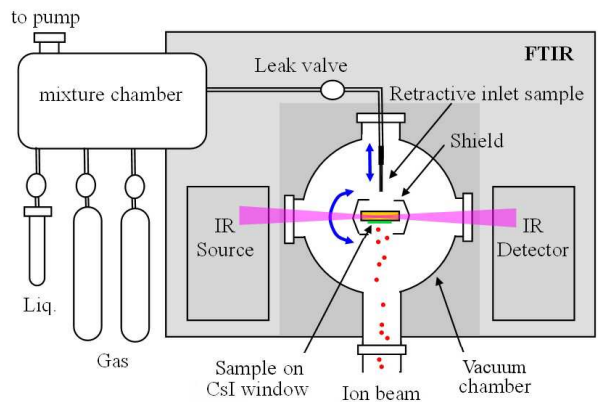}
\caption{Schematic diagram of the experimental set-up. The ion beam impinges perpendicularly on the thin ice film deposited on a CsI crystal (taken from Pilling et al. 2010a).} \label{fig:diagram}
\end{figure}

In this paper, results for pure c-C$_{6}$H$_{12}$ ice and mixed H$_2$O:NH$_3$:c-C$_6$H$_{12}$ (1:0.3:0.7) ice, both at 13 K are presented. Pure c-C$_{6}$H$_{12}$ ice was bombarded by 219 MeV $^{16}$O$^{7+}$ (13.7 MeV/u). For the mixed ice, the projectile employed was 632 MeV $^{58}$Ni$^{24+}$ (10.9 MeV/u). The equilibrium charge state of O and Ni atoms (independent of the initial charge state) after several collisions with matter (ice) is around 7 and 13, respectively (e.g. Nastasi et al. 1996).

\emph{In-situ} infrared spectra were recorded for ices irradiated at different fluences, up to  $6 \times 10^{13}$ ions cm$^{-2}$ using a Nicolet  Fourier-transformed infrared spectrometer (Magna 550) from 4000 to 600 cm$^{-1}$ with 1 cm$^{-1}$ resolution. A background allowing the absorbance measurements was collected before gas deposition. Experimental details are given elsewhere (Seperuelo Duarte et al. 2009; 2010; Pilling et al. 2010a; 2010b).

\begin{table}
\caption{Infrared absorption coefficients (band strengths) used in the column density calculations for the observed molecules.} \label{tab:A}
\setlength{\tabcolsep}{2pt}
\begin{tabular}{ l l l l r }
\hline
Wavenumber  & Wavelength & Assignment  & Band strength                  & Ref. \\
(cm$^{-1}$) & ($\mu$m)  &             &   (cm molec$^{-1}$)             &      \\
\hline
2340         & 4.27 & CO$_2$ ($\nu_3$)    & 7.6$\times 10^{-17}$        & [1]  \\
$\sim$ 2165  & 4.62 & OCN$^-$ ($\nu_3$)    & $\sim$ 4$\times 10^{-17}$  & [2]  \\
2135         & 4.67 & CO ($\nu_1$)         & 1.1$\times 10^{-17}$       & [3]  \\
1299         & 7.69 & CH$_4$ ($\nu_4$)     & 7$\times 10^{-18}$         & [4]  \\
1095         & 9.13 & NH$_3$ ($\nu_2$; umbrella mode)     & 1.2$\times 10^{-17}$   & [5]  \\
861          & 11.6 & c-C$_6$H$_{12}$ (-C-C- stretch) & 8$\times 10^{-19}$                  & [6] \\
$\sim$ 800   & 12.5 & H$_2$O ($\nu_L$; libration mode)     & 2.6$\times 10^{-17}$           & [6]\\
719          & 13.9 & Alkenes (\tiny -C-H out-of-plane bending\normalsize) & $\sim$8$\times 10^{-18}$  & [7] \\
\hline
\end{tabular}
[1] Gerakines et al. 1995; [2] Average value from d'Hendecourt et al. (1986) and Demyk et al. (1998); [3] Jiang et al. 1975;  [4] Gerakines et al. 2005;  [5] Kerkhof et al. 1999; [6] d'Hendecourt \& Allamandola (1986); [7] Tentative average value for simple alkenes, such as cyclohexene (see details in text).
\end{table}

The sample-cryostat system can be rotated over 180$^\circ$ and fixed at three different positions to allow: i) gas deposition, ii) FTIR measurement, and iii) perpendicular irradiation as shown in Fig.~\ref{fig:diagram} (taken from Pilling et al. 2010a). The thin ice film was prepared by condensation of gases (purity superior to 99\%) onto a CsI substrate attached to a closed-cycle helium cryostat,  cooled to 12-13 K. Experiments without sample bombardment did not show any chemical alteration of ice sample due to vacuum with exception of a small water layering as a function of time. During the experiment the chamber pressure was around $ 2\times 10^{-8}$ mbar. After the irradiation the samples were heated ($\sim$ 2 K min$^{-1}$) to room temperature and some IR spectra was taken during this process. Future experiments will be performed to increase the amount of produced organic residues which will be analyzed further by ex-situ chromatographic techniques.

The molecular column densities of samples were determined from the relation between optical depth $\tau_\nu = \ln(I_0/I)$ and band strength, A (cm molec$^{-1}$), of the respective sample vibrational mode. In this expression, $I_0$ and $I$ are the intensity of light at a specific wavenumber, before and after passing through a sample, respectively. Because the absorbance measured by the FTIR spectrometer is $A_\nu=\log(I_0/I)$, the molecular column density of ice samples is given by
\begin{equation} \label{eq-N}
N= \frac{1}{A} \int \tau_\nu d\nu = \frac{2.3}{A} \int a_\nu d\nu \quad \textrm{[molec cm$^{-2}$],}
\end{equation}
where $a_\nu = \ln(I_0/I)/\ln(10) = \tau_\nu/2.3$.

From these measurements and assuming an average density for the ice samples of about 1 g/cm$^3$, the thickness, the deposition rate and the sample mass were determined (see equations in Pilling et al. 2011). For the pure c-C$_6$H$_{12}$, the thickness was about 1.8 $\mu$m, the deposition rate about 11 $\mu$m/h, and the mass about 100 $\mu$g. For the mixed H$_2$O:NH$_3$:c-C$_6$H$_{12}$ (1:0.3:0.7)), the estimated thickness was 4.5 $\mu$m, the deposition rate $\sim$ 27 $\mu$m/h, and the mass $\sim$ 250 $\mu$g.

Knowing the ice thickness and the initial column density and considering that each monolayer has roughly $N_{ML} \sim 10^{15}$ molec cm$^{-2}$, the thickness of a single monolayer, $d_{ML}$, was estimated by the expression
\begin{equation} \label{eq-dML}
d_{ML}= 10^{4} \frac{d}{N_o/N_{ML}} = 10^{19} \frac{d}{N_o}  \quad \textrm{[\AA],}
\end{equation}
where $d$ is the ice thickness ($\mu$m) and $N_o$ is the initial molecular column density (molec cm$^{-2}$). The monolayer thickness of pure cyclohexane ice and mixed H$_2$O:NH$_3$:c-C$_6$H$_{12}$ (1:0.3:0.7) ice were around 15 and 8 \AA, respectively.

\begin{figure} 
 \centering
  \resizebox{\hsize}{!}{\includegraphics{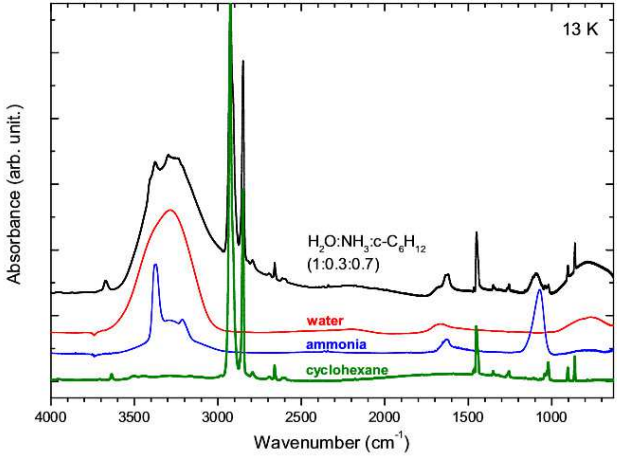}}
\caption{FTIR spectra from 4000-600 cm$^{-1}$ of non-irradiated ices at 13 K: mixed H$_2$O:NH$_3$:c-C$_6$H$_{12}$ (1:0.3:0.7) ice (top), pure c-C$_6$H$_{12}$ ice (bottom). For comparison, spectra of pure H$_2$O ice and pure NH$_3$ ice are also shown.} \label{fig:4ices}
\end{figure}

The analyzed ice layers were thin enough: i) to avoid saturation of the FTIR signal in transmission mode, and ii) to be fully crossed by an ion beam with velocity approximately constant. This latter point is important because the relatively low total kinetic energy loss of the projectile in the film guarantees that the studied cross sections remain constant.

The dissociation cross sections determined from such experiments employing ions have a good reproducibility as we observe for the two H$_2$O:CO$_2$ (1:1) 13 K ices irradiated by 52 MeV Ni ions with the same instrumentation (Pilling et al. 2011). For pure ices, the temperature is the main parameter that can affect experimental reproducibility. However, for mixed ices, besides the temperature,  changes in the initial molecular abundance ratio may affect the entire system, resulting in variations in the dissociation cross sections, formation cross section of newly produced species, as well as, in the sputtering yield (e.g Pilling et al. 2010b).

For comparison, the infrared (IR) spectra of non-bombarded
mixed H$_2$O:NH$_3$:c-C$_6$H$_{12}$ (1:0.3:0.7) ice are presented together with spectra of non-bombarded pure H$_2$O, NH$_3$, and c-C$_6$H$_{12}$ ices in Fig.~\ref{fig:4ices}. Despite the similarities in the IR spectra corresponding to mixed and pure ices, the spectra of mixed ice cannot be obtained by the simple sum of the spectra of pure ices. The presence of different chemical environments in the ice (in the case of mixed ices) changes some IR band profiles and also may change the band strength of some molecular vibration modes. Except for the sharp ammonia  peak at 3400 cm$^{-1}$ (-NH stretching mode), all other sharp peaks in the mixed ices are assigned to c-C$_6$H$_{12}$. The NH$_3$ umbrella vibration mode is easily observed in the mixed ice. However, this band is lightly shifted to higher frequencies (1095 cm$^{-1}$) in comparison with the band´s location observed in pure NH$_3$ ice (1070 cm$^{-1}$). The broad water bands at $\sim$3300 cm$^{-1}$ (-OH stretching mode) and at $\sim$ 800 cm$^{-1}$ (-OH libration mode) are also easily observed in the IR spectra of mixed ices.

The vibrational band positions and infrared absorption coefficients (band strengths) used in this work to derive the column densities of the selected species are given in Table~\ref{tab:A}. Because of the difficulty in determining the area of NH$_3$ and H$_2$O bands due to the convolution with other peaks, their column density values should be employed with caution. In attempt to quantify the amount of produced alkenes, we considered the cleanest and well defined alkene peak in the IR spectra, which occurred at 719 cm$^{-1}$ (13.9 $\mu$m) (=CH out-of-plane bending). For this IR transition, an average band strength value of $\sim$8$\times 10^{-18}$ cm molec$^{-1}$ was adopted. This value is only an approximation for the amount of alkenes, since it represents an average value for this band in simple alkenes, such as hexene, cyclohexene, benzene (e.g. Pavia et al. 2009; Ruiterkamp et al. 2005; d´Hendecourt \& Allamandola, 1986).

\section{Results and discussion} 

\begin{figure} 
 \centering
  \resizebox{\hsize}{!}{\includegraphics{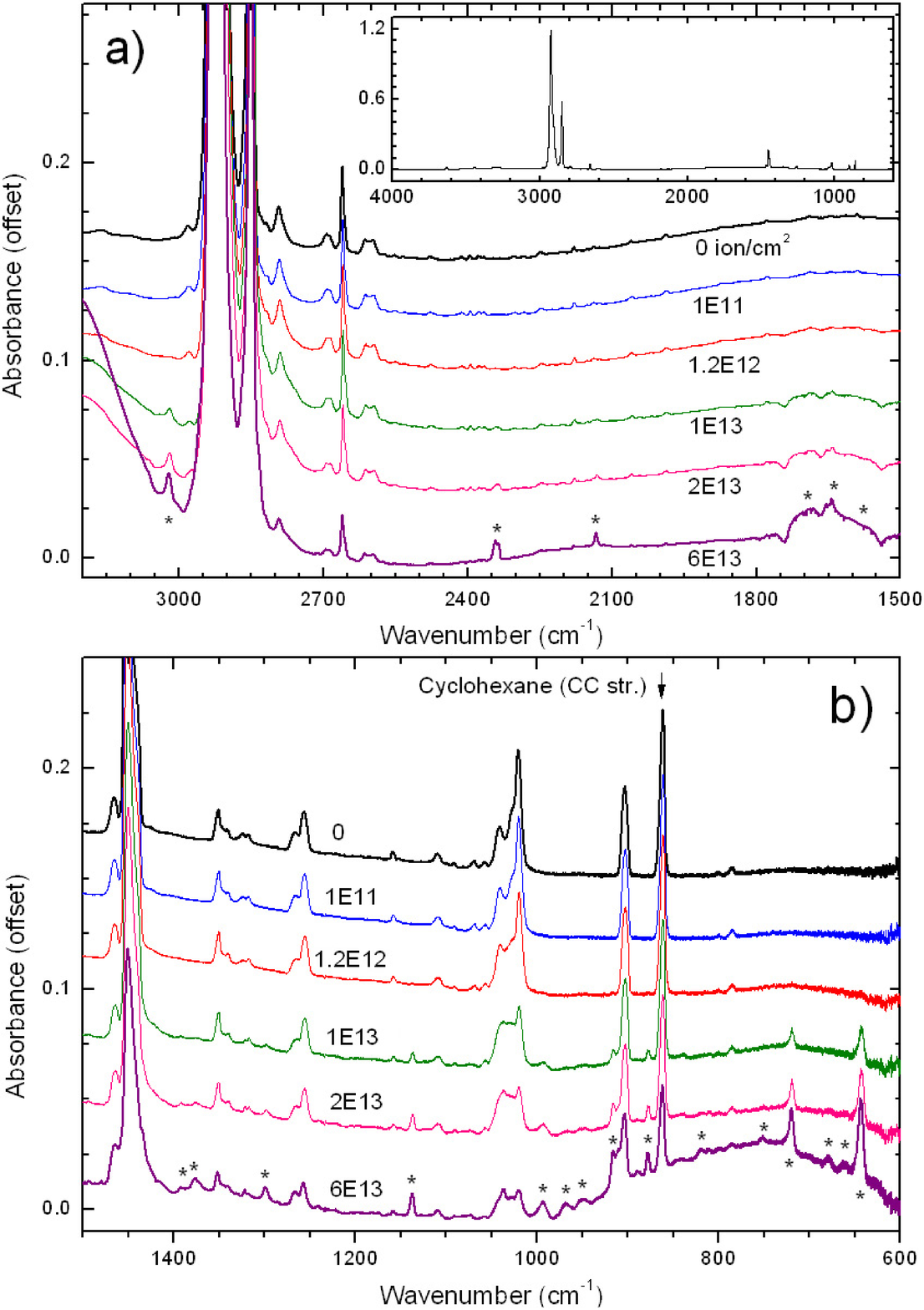}}
\caption{a) Expanded view of the infrared spectra of pure c-C$_6$H$_{12}$ ice at 13 K before (highest curve) and after different irradiation fluences employing 219 MeV O ions (medium-mass cosmic ray analog) from 3200 to 1500 cm$^{-1}$. Figure inset shows IR spectrum of non-bombarded c-C$_6$H$_{12}$ ice from 4000 to 600 cm$^{-1}$. b) Expanded view from 1600 to 600  cm$^{-1}$. Ion fluences are indicated. Asterisks indicate the location of newly produced IR bands.} \label{fig:Pureice}
\end{figure}

Figures~\ref{fig:Pureice}a-b present the evolution of the IR spectra of pure c-C$_6$H$_{12}$ ice at 13 K as a function of ion fluence. In both figures, the uppermost curve shows the IR spectra of pure c-C$_6$H$_{12}$ ice at 13 K before irradiation. Other curves show the IR spectra at different irradiation fluences employing 219 MeV O ions. Each spectrum is offset for clearer visualization. The figure inset shows the IR spectrum of the pure c-C$_6$H$_{12}$ ice as a whole. The infrared absorption band employed to determine the molecular column density of cyclohexane is indicated by the arrow. The lowest curve was obtained after a fluence of $6 \times 10^{13}$ O ions cm$^{-2}$. The residual gas inside the chamber (mainly water) causes the condensation of water (broad bands around 3300 cm$^{-1}$ and 800 cm$^{-1}$), better seen at high fluences, as well as some radiolysis products such as CO (2133 cm$^{-1}$) and CO$_2$ (2341 cm$^{-1}$). In both figures, the newly formed products from the radiolysis of c-C$_6$H$_{12}$ are indicated by asterisks and their possible assignments are listed in Table~\ref{tab:newspec}.

\begin{figure} 
 \centering
 \resizebox{\hsize}{!}{\includegraphics{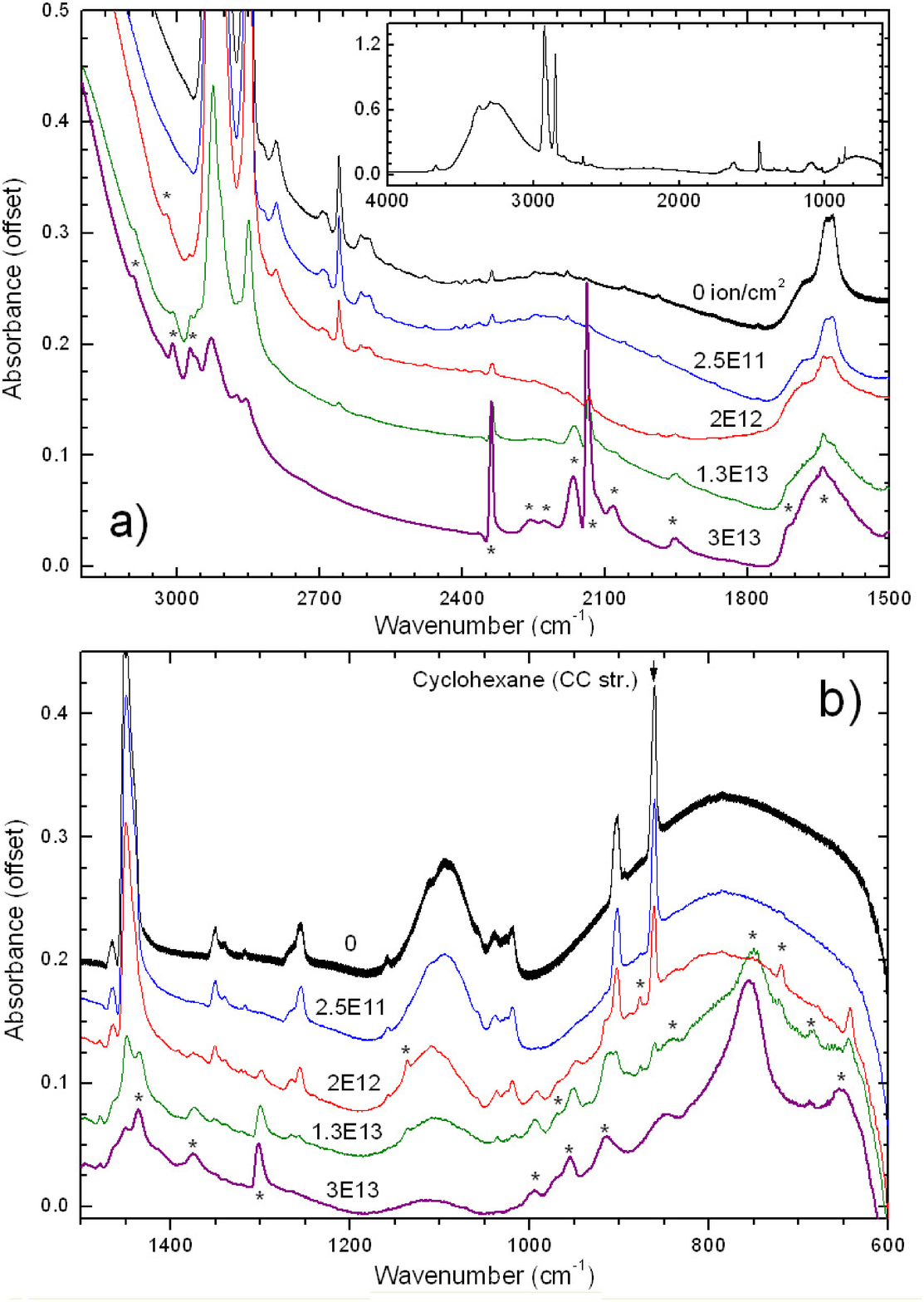}}
\caption{a) Expanded view of the infrared spectra of mixed H$_2$O:NH$_3$:c-C$_6$H$_{12}$ (1:0.3:0.7) ice at 13 K before (highest curve) and after different irradiation fluences employing 632 MeV Ni ions (heavy cosmic ray analog) from 3200 to 1500 cm$^{-1}$. Figure inset shows IR spectrum of non-bombarded mixed ice from 4000 to 600 cm$^{-1}$. b) Expanded view from 1600 to 600 cm$^{-1}$. Ion fluences are indicated. Asterisks indicate the location of newly produced IR bands.} \label{fig:mixedice}
\end{figure}

Figures~\ref{fig:mixedice}a-b present the infrared spectra of mixed H$_2$O:NH$_3$:c-C$_6$H$_{12}$ (1:0.3:0.7) ice at 13 K before (uppermost curve) and after different irradiation fluences employing 632 MeV Ni ions. Each spectrum is offset for clearer visualization. As in the previous figure, the infrared absorption band employed to determine the molecular column density of cyclohexane is indicated by the arrow. The lowest curve was obtained after a fluence of $3 \times 10^{13}$ Ni ions cm$^{-2}$. Figure inset shows IR spectrum of non-bombarded c-C$_6$H$_{12}$ ice from 4000 to 600 cm$^{-1}$. The broad structure from 3100 to 3500 cm$^{-1}$ represents a combination of vibration modes of water ($\nu_1$) and ammonia ($\nu_1$). The narrow peak at 2100 cm$^{-1}$ is the CO stretching mode ($\nu_1$).  The band at 1600 cm$^{-1}$ consists of two lines corresponding to the water $\nu_2$ vibration mode (1650 cm$^{-1}$) and ammonia $\nu_4$ vibration mode (1630 cm$^{-1}$).  The feature around 1100  cm$^{-1}$ is the umbrella vibration ($\nu_2$) mode of ammonia and the one at 800 cm$^{-1}$ is the libration mode ($\nu_L$) of water molecules. The newly formed products from the radiolysis of mixed H$_2$O:NH$_3$:c-C$_6$H$_{12}$ (1:0.3:0.7) ice at 13 K are indicated by asterisks and their assignments are listed in Table~\ref{tab:newspec}.

The formation of CO$_2$ (2341 cm$^{-1}$) from the bombardment of pure c-C$_6$H$_{12}$ is sensitively triggered after employing 10$^{13}$ ions cm$^{-1}$ in the ice. In addition, for higher fluences, this CO$_2$ peak has a shoulder in the lower wavenumber region (2334 cm$^{-1}$) which may be attributed to the -C$\equiv$C- symmetric stretch of non symmetric alkynes (Pavia et al. 2009).

\begin{figure*} 
 \centering
 \includegraphics[scale=0.65]{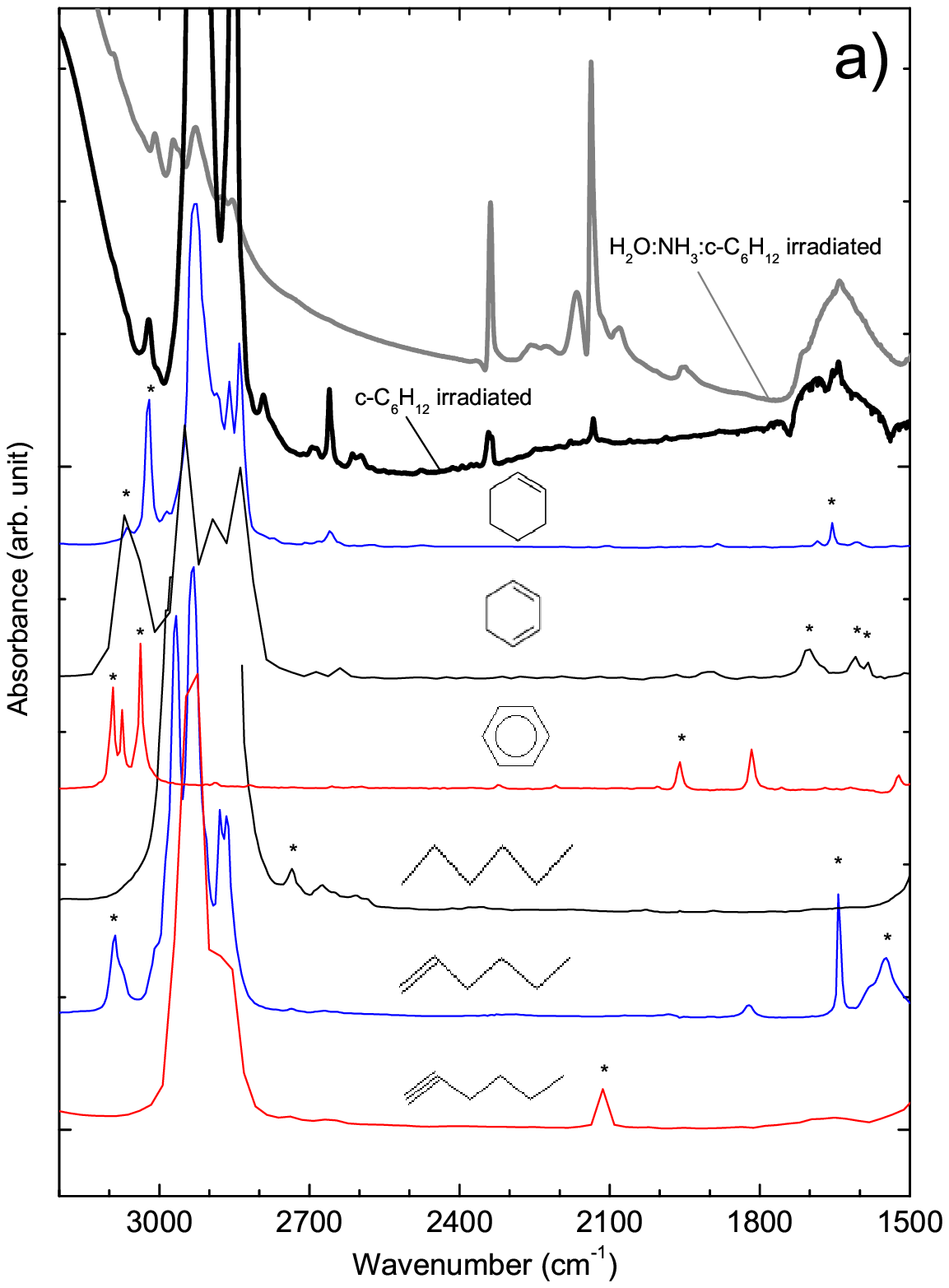}
 \includegraphics[scale=0.65]{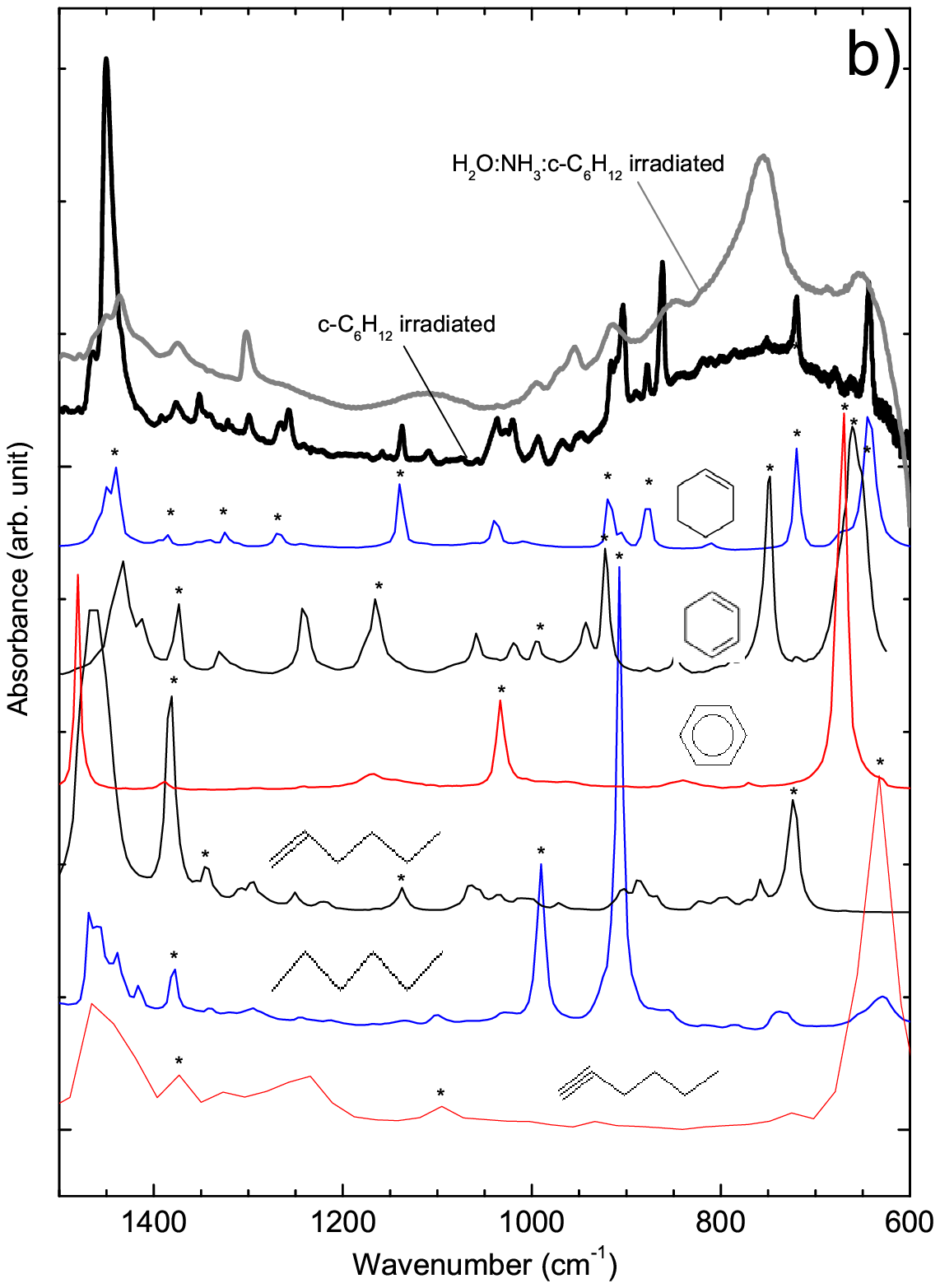}
\caption{Comparison between the IR spectra of the two irradiated ices at highest fluences (this work) with IR spectra of different non irradiated cyclic and aliphatic hydrocarbons from the NIST database. Skeletal formulas of each species (cyclohexene, 1-3 cyclohexadiene, benzene, hexane, 1-hexene, 1-hexyne) are shown. Asterisks indicate the peaks that have possible identification in the spectra of irradiated ices. a) From 3200 to 1500 cm$^{-1}$. b) From 1600 to 600 cm$^{-1}$.} \label{fig:compNIST}
\end{figure*}

Figures~\ref{fig:compNIST}a-b present a comparison of the IR spectra of the two irradiated ices (at highest fluences) in this study with the IR spectra of different non irradiated cyclic and aliphatic hydrocarbons from NIST database\footnote{\texttt{http://webbook.nist.gov/chemistry/}} (liquid-phase). Skeletal formulas of each species (cyclohexene, 1-3 cyclohexadiene, benzene, hexane, 1-hexene, and 1-hexyne) are also shown. Asterisks indicate the peaks that have possible identification in the spectra of irradiated ices. Figure~\ref{fig:compNIST}a shows the IR spectra from 3200-1500 cm$^{-1}$ and Fig.~\ref{fig:compNIST}b presents the 1500-600 cm$^{-1}$ wavenumber range. The presence of a broad structure at 1600 cm$^{-1}$ indicates the presence of the rings of aromatic systems in the ice after bombardment as result of dehydrogenation of cyclohexane molecules. A theoretical study about the energies and the transient states involved during the ring opening of cyclohexane to produce 1-hexene (c-C$_6$H$_{12}$ $\rightarrow$ 1-C$_6$H$_{12}$) was performed by Sirjean et al. (2006). The authors also suggested that species such as birradicals ($\bullet$C$_6$H$_{12}\bullet$ and $\bullet$C$_4$H$_{8}\bullet$) and ethylene (C$_2$H$_4$) are present in such ring open mechanisms. This last species is required in the production of aromatic compounds in interstellar environments (e.g. Frenklach \& Feigelson 1989).

Table~\ref{tab:newspec} lists the new IR bands formed after ion bombardment of the cyclohexane containing ices at 13 K. A large number of vibrational modes are attributed to alkenes and possibly to ring/aromatic systems. For the mixed  H$_2$O:NH$_3$:c-C$_6$H$_{12}$ (1:0.3:0.7) ice, the formation of CO, CO$_2$, and OCN$^-$ are observed. Due to the local heat and energy delivered by projectiles in the sample, the bombardment is expected to promote desorption of molecular hydrogen, however no measurement of this species has been performed.

\subsection{Cross section and radiolysis yield}

The column density variation as a function of ion fluence for each condensed molecular species (parental or daughter) during radiolysis can be written as
\begin{equation} \label{eq:PAI}
\frac{dN_i}{dF} = \sum_{k \neq i} \sigma_{f,ik} N_k + L_i  - \sigma_{d,i} N_i - Y_i \Omega_i(F)
\end{equation}
where $\sum_k \sigma_{f,ik} N_k$ represents the total molecular production rate of the $i$ species directly from the $k$ species, $L_i$ is the layering, $\sigma_{d,i}$ the dissociation cross section, $Y_i$ the sputtering yield, and $\Omega_i(F)$ the relative area covered by the $i$ species on the ice surface. For more details see equations in Pilling et al. (2010a) and de Barros et al. (2011).

For pure cyclohexane ice, the solution of Eq.~\ref{eq:PAI}, considering no extra production($\sigma_{f,ik}=0$), full covering ($\Omega_i(F)=1$) and, no layering (L=0), is given analytically by
\begin{equation}\label{eq:pure}
N= (N_{0} + \frac{Y}{\sigma_{d}}) \exp(-\sigma_{d} F) - \frac{Y}{\sigma_{d}}
\end{equation}
where $N_{0}$, $Y$, and $\sigma_{d}$ are the initial column density, sputtering and dissociation cross section of cyclohexane species, respectively.

In the case of mixed H$_2$O:NH$_3$:c-C$_6$H$_{12}$ (1:0.3:0.7) ice, after the rapid ice compaction phase at the beginning of the irradiation, water layering tends to recover the ice surface ($\Omega_i(F) \rightarrow 0$) progressively preventing the sputtering of other species. Therefore, under the assumption that both sputtering and layering are negligible (Y=L=0) the solution of Eq.~\ref{eq:PAI} for the column density of c-C$_6$H$_12$ as a function of fluence is obtained analytically by
\begin{equation}\label{eq:mixed}
N = N_{0} \exp(-\sigma_{d} F)
\end{equation}
where $N_{0}$, and $\sigma_{d}$ are the initial column density, and dissociation cross section of cyclohexane species in the mixed ice, respectively.

For water (and viscous liquids), it may occur  an additional layering due to the molecular retention in the chamber walls. In this case, and considering no other formation source ($\sigma_{f,ik}=0$), the column density evolution with the fluence is obtained directly by solving Eq.~\ref{eq:PAI}, which gives
\begin{equation}\label{eq:NH2O}
N= (N_{0} - N_{\infty}) \exp(-\sigma_{d} F) + N_{\infty} 
\end{equation}
where $N_{0}$, and $\sigma_{d}$ are the initial column density and dissociation cross section of H$_2$O, respectively. In this equation, $N_{\infty} = (L - Y)/ \sigma_{d}$ is the asymptotic value of column density of water due to the presence of layering. In experiments where high layering values are observed, the sputtering may be considered negligible.

As discussed by de Barros et al. (2011), for low fluences (up to $\sim 5 \times 10^{12}$ ions cm$^{-2}$), the column density evolution of daughter species produced by the radiolysis of both pure and mixed ices is
\begin{equation}\label{eq:daughter}
N_k \approx N_{0} \sigma_{f,k}  \big( F - \frac{\sigma_{d} + \sigma_{d,k} }{2}F^2 \big)
\end{equation}
where $N_k$ is the column density of daughter species $k$ at a given fluence. $\sigma_{f,k}$ and  $\sigma_{d,k}$ are the formation cross section and dissociation cross section of daughter species $k$, respectively. $N_{0}$ and $\sigma_{d}$ are the initial column density and dissociation cross section of parental species, respectively. This equation assumes a negligible value for the sputtering of daughter species ($Y_k$=0).

\begin{table}
\small
\caption{Stopping power and penetration depth values for 219 MeV (13.6 MeV/u) $^{16}$O ions and 632 MeV (10.8 MeV/u) $^{58}$Ni ions in the studied ices, calculated by the SRIM code. For comparison, values for 632 MeV (10.8 MeV/u) $^{58}$Ni, 1 MeV protons, and 12 MeV protons all in pure cyclohexane ice, and 1 MeV protons in pure water ice are also shown.} \label{tab:dedxSRIM}
\setlength{\tabcolsep}{3pt}
\begin{tabular}{ l c c c c }
\hline
Projectile     & \multicolumn{3}{c}{Stopping power}   & Penetration    \\
        \cline{2-4}
        & Electronic      & Nuclear   & Total   & depth ($\mu$m)           \\
        & (keV$/\mu$m)   & (keV$/\mu$m)   & (eV/(molec/cm$^2$)) &      \\
\hline
219  MeV O$^{a}$  & 250.4  & 0.13  & 3.49 $\times 10^{-13}$   & 509.7   \\
632  MeV Ni$^{b}$ & 2787   & 1.70  & 1.27 $\times 10^{-12}$   & 172.5   \\
\hline
632  MeV Ni$^{a}$ & 2802  & 1.72   & 3.88 $\times 10^{-12}$   & 169.8   \\
1   MeV p$^{a}$   & 28.89 & 0.023  & 3.12 $\times 10^{-15}$   & 21.4   \\
1   MeV p$^{c}$   & 25.11 & 0.021  & 7.51 $\times 10^{-13}$   & 25.1   \\
12  MeV p$^{a}$   & 4.296 & 0.002  & 4.03 $\times 10^{-14}$   & 1540   \\
\hline
\end{tabular}
Employed ice density = 1.0  g cm$^{-3}$ .\\
$^{a}$ in pure c-C$_6$H$_{12}$ ice ($\sim$ 7 $\times 10^{21}$ molec cm$^{-3}$). \\
$^{b}$ in mixed H$_2$O:NH$_3$:c-C$_6$H$_{12}$ (1:0.3:0.7) ice ($\sim$ 2 $\times 10^{22}$ molec cm$^{-3}$). \\
$^{c}$ in pure H$_2$O ice ($\sim$ 3 $\times 10^{22}$ molec cm$^{-3}$). \\
\end{table}  %

\begin{figure}
 \centering
\resizebox{\hsize}{!}{\includegraphics{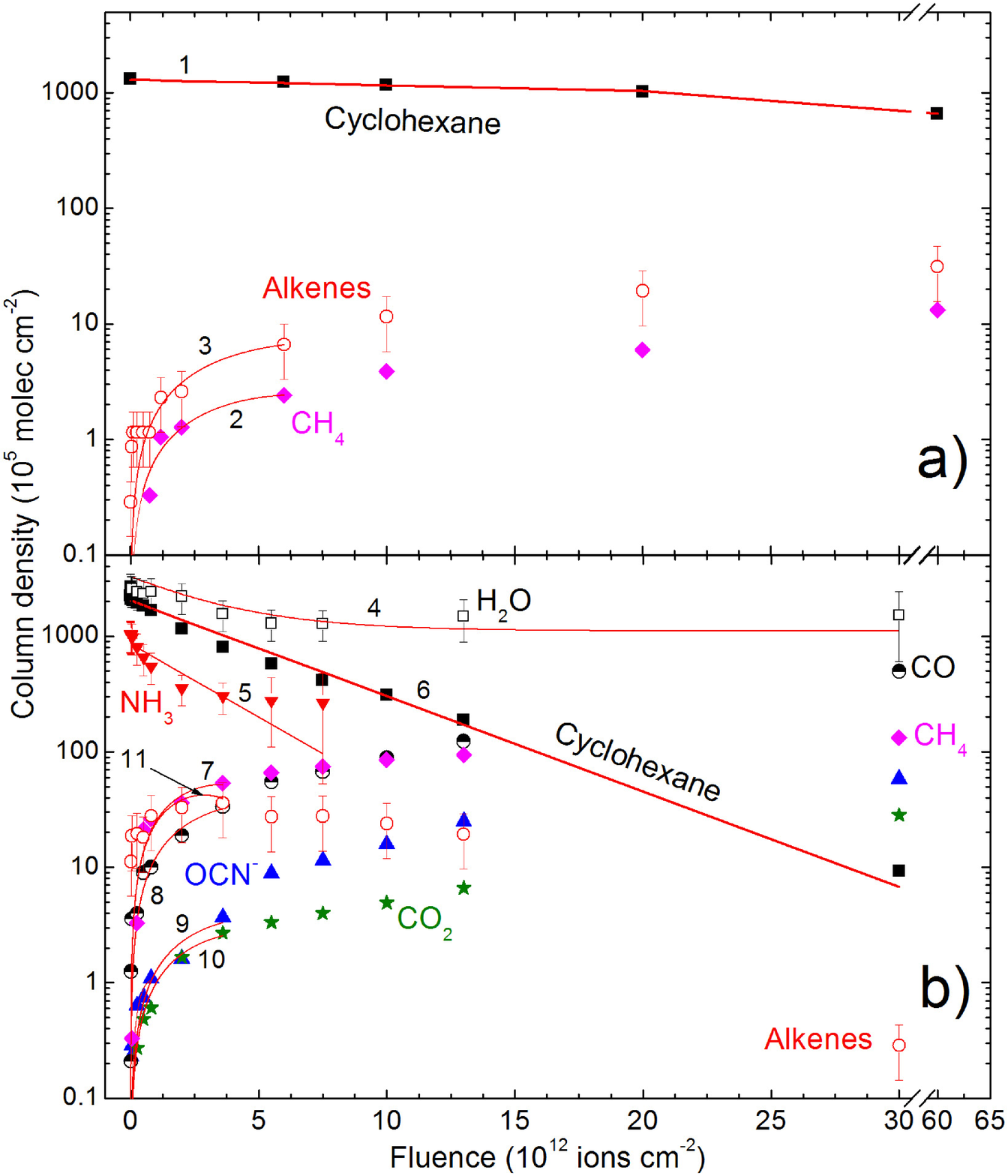}}
\caption{a) Variation of the column density of cyclohexane and selected daughter species (CH$_4$) in pure ice experiment as function of ion (219 MeV O) fluence. b) Variation of the column density of cyclohexane, water and ammonia and selected daughter species (CO, CO$_2$, CH$_4$ and OCN$^-$) in mixed ice experiment as a function of ion (632 MeV Ni) fluence. The lines indicate the fittings using Eq.~\ref{eq:pure} (pure cyclohexane), Eq.~\ref{eq:mixed} (cyclohexane and ammonia in mixed ice), Eq.~\ref{eq:NH2O} (water in mixed ice) and, Eq.~\ref{eq:daughter} (products). The model parameters are given in Table~\ref{tbl:models}.}
\label{fig:column}
\end{figure}

Figure~\ref{fig:column} shows the evolution of the column density of cyclohexane in pure and mixed studied ices as a function of ion fluences. The column densities of four radiolysis products (CO, CO$_2$, CH$_4$, and OCN$^-$) observed in the radiolysis of the mixed ice are also shown. The decreasing of the cyclohexane column density is related to the formation of other species and to the sputtering induced by heavy-ions (Seperuelo Duarte et al. 2009; 2010; Pilling et al. 2010a; 2010b).

Table~\ref{tab:dedxSRIM} shows the stopping power and penetration depth values for 219 MeV (13.6 MeV/u) $^{16}$O ions and 632 MeV (10.8 MeV/u) $^{58}$Ni ions in the studied ices, calculated by the Stopping and Range of Ions in Matter - SRIM code\footnote{\texttt{http://www.srim.org/}}. The SRIM code is a collection of software packages that calculate many features of the transport of ions in matter (Ziegler et al. 2011). For comparison, values for 632 MeV (10.8 MeV/u) $^{58}$Ni, 1 MeV protons and 12 MeV protons, all in pure cyclohexane ice, and 1 MeV protons in pure water ice are also shown. The energy delivered per micron by a single 632 MeV Ni ion is roughly 11 times higher than for a single 219 MeV O ion, and about 110 times higher than for a single 1 MeV proton.

The radiochemical formation yield ($G_f$) of a given compound per 100 eV of deposited energy, at normal incidence, is written by
\begin{equation} \label{radiochemical}
G_f = 100 \frac{\sigma_f}{S}   \quad \textrm{molecules per 100 eV.}
\end{equation}
where $\sigma_f$ is the formation cross section and \emph{S} is the stopping power, in units of eV/(molec/cm$^2$) (Loeffler et al. (2005).

\begin{table*}
\caption{Cross sections ($\sigma_f$ and  $\sigma_d$), radiochemical yield (G$_f$ and G$_d$), and sputtering yield (Y) for cyclohexane (and H$_2$O, NH$_3$ for the mixed ice) and selected daughter species obtained from ion bombardment experiments of 13 K ices. The layering (L) and the initial column density (N$_0$) are also listed. The fittings are shown in Fig.~\ref{fig:column}.}
 \label{tbl:models}
  \setlength{\tabcolsep}{5pt}  
\begin{tabular}{l c c c c c c c c}
\multicolumn{9}{l}{\textbf{Pure c-C$_6$H$_{12}$ ice irradiated with 219 MeV O$^{7+}$}} \\
\hline
species$^a$  & $\sigma_f$          & $\sigma_d$            & G$_f$                                       & -G$_d$                                      & $Y$                                             & $L$                                            &   $N_0$                                             & Model    \\
             & $(10^{-13}$ cm$^2$) & $(10^{-13}$ cm$^2$)   & $(\frac{\textrm{molec}}{\textrm{100 eV}})$  & $(\frac{\textrm{molec}}{\textrm{100 eV}})$  & (10$^4$ $\frac{\textrm{molec}}{\textrm{ion}}$) & (10$^4$ $\frac{\textrm{molec}}{\textrm{ion}}$) &  (10$^{18}$ $\frac{\textrm{molec}}{\textrm{cm}^2}$)   &            \\
\hline
c-C$_6$H$_{12}$ &  0$^a$      & 0.1          & 0$^a$       & 2.9      & 0.1  & 0 & 1.3      & 1  \\
CH$_{4}$        &  0.006      & 1.4          & 0.17        & 40      & 0    & 0 & 0       & 2  \\
Alkenes$^c$     &  $\sim$0.01 & $\sim$ 1.2   & $\sim$ 0.3  & $\sim$ 34 & 0    & 0 & 0       & 3  \\
\\
\\
\\
\multicolumn{9}{l}{\textbf{Mixed H$_2$O:NH$_3$:c-C$_6$H$_{12}$ (1:0.3:0.7) ice irradiated with 632 MeV Ni$^{24+}$}} \\
\hline
species$^a$      & $\sigma_f$          & $\sigma_d$            & G$_f$                                       & -G$_d$                                      & $Y$                                             & $L$                                            &   $N_0$                                             & Model    \\
             & $(10^{-13}$ cm$^2$) & $(10^{-13}$ cm$^2$)   & $(\frac{\textrm{molec}}{\textrm{100 eV}})$  & $(\frac{\textrm{molec}}{\textrm{100 eV}})$  & (10$^4$ $\frac{\textrm{molec}}{\textrm{ion}}$)  & (10$^4$ $\frac{\textrm{molec}}{\textrm{ion}}$)  &  (10$^{18}$ $\frac{\textrm{molec}}{\textrm{cm}^2}$)   &            \\
\hline
H$_2$O         &  0$^a$     & $\sim 3$  & 0$^b$  & $\sim$ 24  & 1$^d$   & 300  & 3.3     & 4  \\
NH$_3$         &  0$^a$     & $\sim 3$  & 0$^b$  & $\sim$ 24  & 0   & 0    & 0.86    & 5  \\
c-C$_6$H$_{12}$ &  0$^a$    & $\sim 2$  & 0$^b$  & $\sim$ 16  & 0   & 0    & 2.0     & 6  \\
CH$_{4}$       &  0.14      & 0.9       & 1.1     & 7.1         & 0    & 0 & 0       & 7   \\
CO             &  0.07      & $<$ 0.01  & 0.55    & $<$ 0.08  & 0   & 0    & 0      & 8  \\
OCN$^-$        &  0.007     & $<$ 0.01  & 0.06    & $<$ 0.08  & 0   & 0    & 0      & 9  \\
CO$_2$         &  0.006     & $<$ 0.1   & 0.05    & $<$ 0.8  & 0    & 0    & 0      & 10  \\
Alkenes$^c$    &  $\sim$0.1 & $\sim$ 1.5  & $\sim$ 0.8  & $\sim$ 12  & 0    & 0 & 0                         & 11  \\

\hline
\multicolumn{9}{l}{$^a$ Band position given in Table~\ref{tab:A}.}\\
\multicolumn{9}{l}{$^b$ Considering a negligible production of parental species during the radiolysis.}\\
\multicolumn{9}{l}{$^c$ Considering the C-H out-of-plane bending mode at 719 cm$^{-1}$ ($A \sim 8\times 10^{-18}$ cm molec$^{-1}$).}\\
\multicolumn{9}{l}{$^d$ Taken from Brown et al. 1984}\\
\end{tabular}
\end{table*}

This definition can be extended to the radiochemical dissociation (destruction) yield of a given compound per 100 eV of deposited energy ($G_d$) by replacing the formation cross section in Eq.~\ref{radiochemical} with the negative value of the dissociation cross section (-$\sigma_d$). Therefore, negative $G_d$ values indicate that molecules are being dissociated or destroyed after energy deposition into the ice. By adopting the $S$ values from the stopping and ranges module of the SRIM code, the values of the radiation yield $G$ in the experiments can be determined and compared with values in the literature.

Figures~\ref{fig:column}a-b present the column density evolution of species initially present in pure and mixed ice experiments, as well some newly formed products due to the radiolysis, as a function of ion fluences. Fig.~\ref{fig:column}a shows the evolution of cyclohexane, the newly formed CH$_4$, as well as the possible alkenes produced from the bombardment of the pure cyclohexane ice by 219 MeV O ions. The evolution of cyclohexane, water and ammonia, four newly formed species (CO, CO$_2$, CH$_4$ and OCN$^-$), and possible alkenes observed in the radiolysis of the mixed ice by 632 MeV Ni ions are given in Fig.~\ref{fig:column}b. The lines indicate the fittings using Eq.~\ref{eq:pure} (pure cyclohexane), Eq.~\ref{eq:mixed} ( cyclohexane and ammonia in mixed ice), Eq.~\ref{eq:NH2O} (water in mixed ice) and, Eq.~\ref{eq:daughter} (products). Numeric labels indicate the models employed in which the parameters are listed in Table~\ref{tbl:models}.

Table~\ref{tbl:models} list the values obtained by best-fitting curve employing Eqs.~\ref{eq:pure}, \ref{eq:mixed}, \ref{eq:NH2O} and, \ref{eq:daughter} to the column density evolution data of bombarded ices. First columns show the selected species, followed by the formation cross section (only for daughter species), dissociation cross section and, radiolysis yields (formation and destruction). For the radiolysis yields we employed the stopping power values listed in Table~\ref{tab:dedxSRIM}, $3.49 \times 10^{-13}$ eV/(molec/cm$^2$) (for pure ice) and  $1.27 \times 10^{-12}$ eV/(molec/cm$^2$) (for mixed ice). In addition, for mixed ice it was considered that the projectile energy was absorbed by the whole bulk sample and not only for a single species. The last three columns in Table~\ref{tbl:models} are the sputtering yield (only for pure ice), the layering and the initial column density. The sputtering value for water of 10$^4$ molec per impact was taken from Brown et al. (1984). The water layering was derived by the equation $L = N_\infty  \sigma_d + Y$.

The destruction cross section obtained in this study (around 10$^{-13}$ cm$^{2}$) for the frozen species bombarded with 632 MeV Ni ions are in the same order of magnitude of the values obtained previously also employing swift heavy-ions (Pilling et al. 2010a; 2010b; 2011). However, the formation cross sections and also the formation yields of new species decrease with the projectile energy. This issue reinforces that an optimal energy associated with the formation of new (organic) species during the processing of astrophysical ices could exist (see also Pilling et al. 2011).

The formation cross section of alkenes, considering the C-H out-of-plane bending mode at 719 cm$^{-1}$ ($A \sim 8\times 10^{-18}$ cm molec$^{-1}$), in the pure c-C$_{6}$H$_{12}$ ice and mixed H$_2$O:NH$_3$:c-C$_6$H$_{12}$ ice were about $1\times 10^{-15}$ and $1\times 10^{-14}$ cm$^{2}$, respectively. These can be compared with similar experiments employing other saturated hydrocarbons such as ethane, propane and cyclopropane. Future experiments will compare the formation cross section of alkenes (-C=C-), induced by cosmic rays in interstellar ice analogs, with respect to the number of carbons of parental saturated hydrocarbon present initially in the ice.

\section{Astrophysical implications} 

\begin{figure}
 \centering
\resizebox{\hsize}{!}{\includegraphics{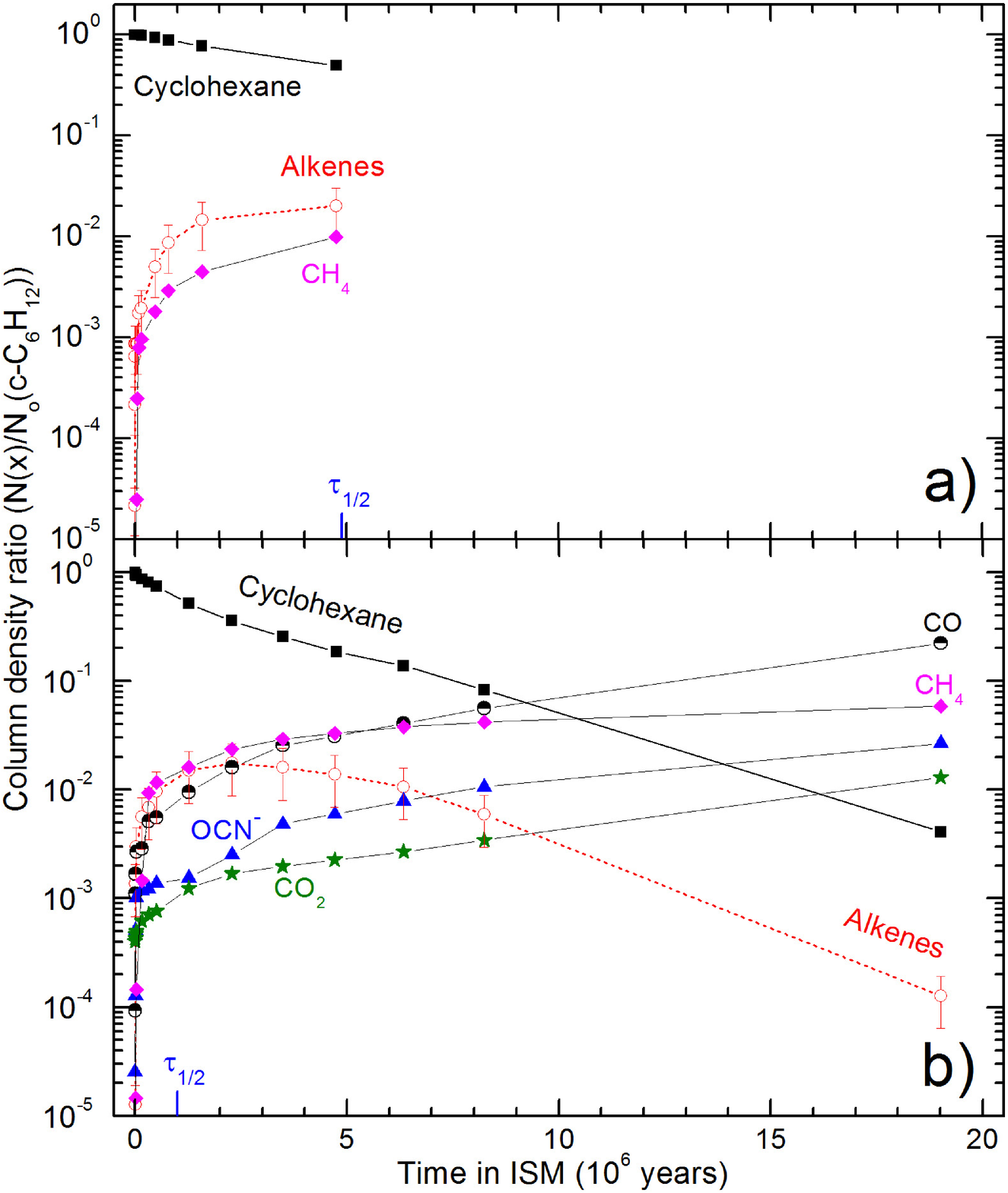}}
\caption{Column density ratios of cyclohexane and some radiolysis products over initial cyclohexane column density as function of equivalent cosmic ray exposure time at ISM. a) pure c-C$_6$H$_{12}$ ice at 13 K irradiated by 219 MeV O ions (medium-mass cosmic ray analog). b) mixed H$_2$O:NH$_3$:c-C$_6$H$_{12}$ (1:0.3:0.7) ice at 13 K irradiated by 632 MeV Ni ions (heavy cosmic ray analog). In each figure, the half-life of cyclohexane ($t_{1/2}$) in ISM as a result of cosmic ray bombardment is indicated. Lines are only to guide the eyes.}
\label{fig:timeISM}
\end{figure}

Despite the ion flux employed in laboratory experiments be several orders of magnitude higher than the flux of similar ions in space and, the thickness of laboratory ice be also higher, the energy delivered and damage induced by similar ions in both scenarios are similar. For example, lets quantify the role of ice size in the experiments. As discussed in section 2, the thickness of pure cyclohexane ice was about 1.5 $\mu$m and, the stopping power of the employed projectile in this ice was 250 keV $\mu$m$^{-1}$ (see Table 2). Thus, the energy difference between the projectile at the ice surface (first layer) and after goes through the ice was 0,375 MeV ($\sim$0,2\% of projectile energy). For the mixed ice, the thickness was around 4.5 $\mu$m and the projectile stopping power was 2788 keV $\mu$m$^{-1}$, resulting in the projectile´s energy difference, between top and bottom ice layers, of about 12,5 MeV ($\sim$2\% projectile energy). Therefore, in both cases we consider a constant projectile energy within the sample and also a linear energy deposit. These assumption are in a good agreement with cosmic ray impact in submicron ices in interstellar or interplanetary regions.

In a similar experiment performed by Pilling et al. (2010a), the authors discussed that each projectile induces significant changes in the sample only in a region with roughly 3 nm of diameter. Considering a constant and homogeneous ion flux of $2\times 10^9$~ions~cm$^{-2}$~s$^{-1}$, the average distance between two nearby impacts is roughly 300 nm (about one hundred times higher the length of processed sample by a single projectile hit). From these values, we estimate that the probability of an ion to hit a given area of 3 nm of diameter in each second is about $P \sim 1.4\times 10^{-4}$. Therefore, only after about two hours of continuous bombardment we expect a second projectile hit in the same nanometric region. This is enough time to consider that, in laboratory, each projectile hit always a thermalized region. Such scenario is very similar to interstellar or interplanetary low flux conditions (over a very extended period of time).

The equivalent cosmic ray exposure time ($T_{ISM}$) in interstellar medium with respect to the experimental ion fluence can be obtained by the equation:
\begin{equation}\label{eq:Time_ISM}
T_{ISM} \simeq 3.2 \times 10^{-8} \frac{F}{\phi} \quad \quad \textrm{[year]}
\end{equation}
where $F$ is the ion fluence employed in the experiments, in units of ions cm$^{2}$, and $\phi$ indicates the flux of galactic cosmic rays (e.g. for medium mass cosmic rays $\phi_{MCR}$ or for heavy cosmic rays $\phi_{HCR}$) in units of ions cm$^{-2}$ s$^{-1}$. Pilling et al. (2010a) estimated the integrated cosmic ray flux for heavy-ion (12~$\lesssim$~Z~$\lesssim$~29) component with energies between 0.1-10 MeV/u in interstellar medium. The value found was $\phi_{HCR} \sim 5 \times 10^{-2}$~cm$^{-2}$~s$^{-1}$. Pilling et al 2011, employing a similar methodology, estimated the medium-mass (3~$\lesssim$~Z~$\lesssim$~11) component of galactic cosmic rays in interstellar medium with the same energy range. The value obtained by those authors was $\phi_{MCR} \sim 4 \times 10^{-1}$~cm$^{-2}$~s$^{-1}$. Following those authors, both values are also a good estimative for the heavy-ion and medium-mass components of galactic cosmic rays at the outer border of the heliopause.

The chemical evolution of the studied ices extrapolated to space conditions (dust grains in interstellar medium) is shown in Fig.~\ref{fig:timeISM}a-b. In this figure, lines are only to guide the eyes. The column density ratios of cyclohexane and some of its radiolysis products over initial cyclohexane column density, as function of equivalent exposure time in ISM by cosmic rays, are shown in Fig.~\ref{fig:timeISM}a for pure c-C$_6$H$_{12}$ ice at 13 K irradiated by 219 MeV O ions (medium-mass cosmic ray analog), and in Fig.~\ref{fig:timeISM}b for H$_2$O:NH$_3$:c-C$_6$H$_{12}$ (1:0.3:0.7) ice at 13 K irradiated by 632 MeV Ni ions (heavy cosmic ray analog). The total column density of produced alkenes was obtained by using the area of the IR feature at 719 cm$^{-1}$ (-CH bending out plane) considering, as a first hypothesis, that this band can be representative of all the alkenes in the ice. This value represents an intermediate value between the strongest and weakest IR features in the case of clyclohexane (e.g. d´Hendecourt \& Allamandola, 1986). The error bars in the figure reflect the uncertainty in the methodology employed to derive the column density of alkenes from IR spectra. The maximum production of alkenes is obtained after about 3-5 $\times 10^6$ years, independently of the projectile type. Its maximum value was roughly the same in both experiments, $\sim$ 10$^{-2}$ alkenes per cyclohexane molecule. However, this value must be adopted with caution since to the employed assumption for the column density of alkenes may not be representative for all alkenes. Better determination of the band strengths of alkenes for the 719 cm$^{-1}$ IR feature in ice phase is needed to make this methodology more accurate.

\begin{figure*}
 \centering
\includegraphics[scale=0.29]{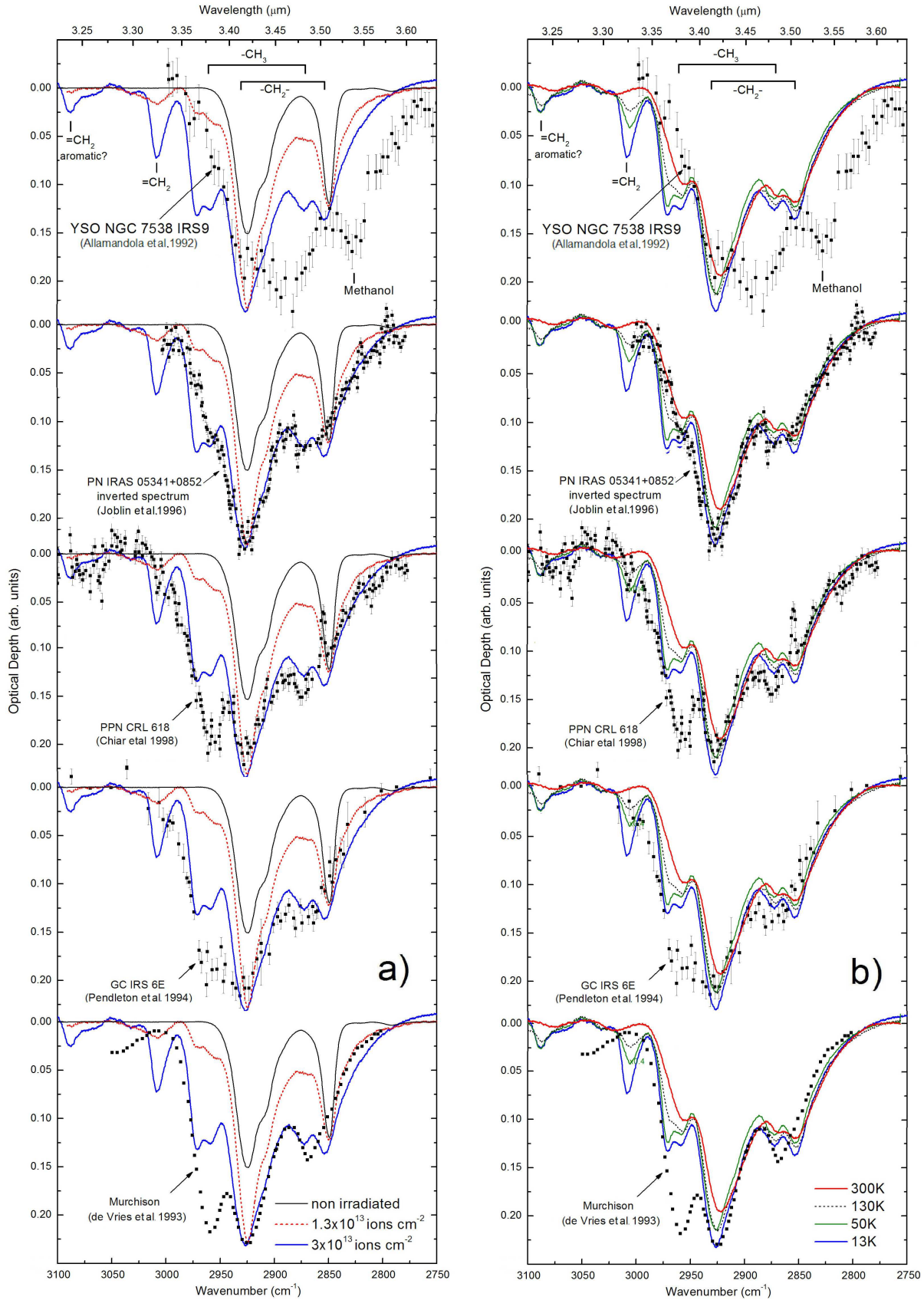}
\caption{Comparison between the 3.4 $\mu$m band of four different astronomical observations (Young Stellar Object NGC 7538 IRS9 (Allamandola et al. 1992); Planetary Nebula IRAS 05341+0852 (Joblin et al. 1996); Protoplanetary Nebula CRL 618 (Chiar et al. 1998); Diffuse interstellar medium at galactic center IRS 6E (Pendleton et al. 1994), and the Murchison meteorite extract (de Vries et al. 1993) with the FTIR spectra of H$_2$O:NH$_3$:c-C$_6$H$_{12}$ ice (this work) at: a) three different ion fluences; b) during sample heating from 13 K up to room temperature.} \label{fig:comp3p4micron}
\end{figure*}

Employing the estimated cosmic ray ion flux and the determined dissociation cross section, the half-life of cyclohexane ($t_{1/2}$) in ISM as a result of cosmic ray bombardment for pure cC$_6$H$_{12}$ ice is about $5 \times 10^6$ years (considering only medium-mass cosmic rays) and for mixed ice is about $1 \times 10^6$ years (considering only heavy-ion cosmic rays). These values are indicated in Fig.~\ref{fig:timeISM}. The results also show that for the mixed ice (H$_2$O:NH$_3$:c-C$_6$H$_{12}$), after 20 $\times 10^6$ years in ISM, almost 20\% of the initial cyclohexane was converted into CO by heavy cosmic rays, 3\% was transformed into OCN$^-$ and 1\% into CO$_2$. This suggests that highly hydrogenated hydrocarbons in water-rich grain mantles inside interstellar clouds can be largely converted into CO during the lifetime of the cloud.

Figure~\ref{fig:timeISM}b also suggests that some daughter species such as CH$_4$ (1300 cm$^{-1}$) and OCN$^-$ (2165 cm$^{-1}$) can be used to estimate the integrated dose of incoming radiation and, assuming a constant cosmic ray flux over the time, the exposure time of interstellar ices to the cosmic rays. At ion fluences higher than 3 $\times$ 10$^{12}$ ions cm$^2$ ($\sim 1 \times 10^{16}$ years in ISM), the abundance of these species increases almost linearly with the fluence.

\subsection{3.4 $\mu$m band}

The 3.4 $\mu$m ($\sim$2925 cm$^{-1}$) band observed in the IR spectra of several astrophysical sources, including protostellar ices, hot grains in diffuse interstellar medium, dust grains in planetary nebulae and meteorites extractions, is frequently associated with hydrocarbons (e.g Allamandola et al. 1992, and references therein). As pointed out by Pendleton and Allamandola (2002), although the 3.4 $\mu$m band probes only the CH stretching modes of the carrier, the analysis of the peak position, and the band profile has revealed that this portion of the refractory material contains aliphatic hydrocarbons (open chain -CH$_2$- and -CH$_3$ groups) that incorporate at least 3\% of the available interstellar carbon. In addition, the other IR features near this band, for example around 3.2-3.3 $\mu$m, also indicate the presence of alkenes and aromatic rings (=CH$_2$ sp$^2$).

\begin{figure}
 \centering
\includegraphics[scale=0.14]{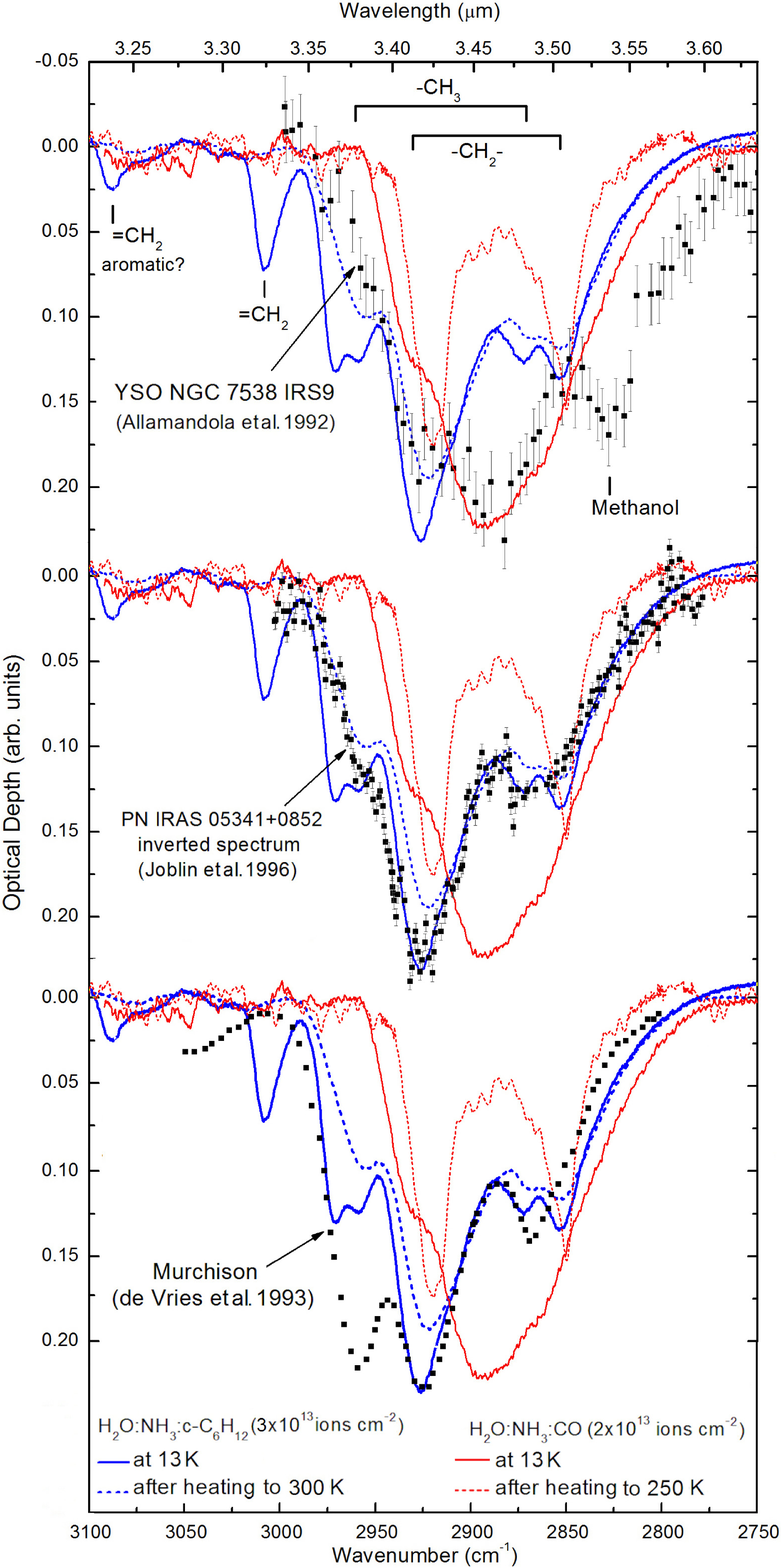}
\caption{Comparison between the radiolysis products from two simulated interstellar ices at 13 K and after warming (H$_2$O:NH$_3$:CO (1:0.6:0.4) bombarded with 46 MeV $^{46}$Ni$^{13+}$ (Pilling et al. 2010a) and H$_2$O:NH$_3$:c-C$_6$H$_{12}$ (1:0.3:0.7) bombarded with 632 MeV $^{58}$Ni$^{24+}$ (this work). Each set in this figure also compares the laboratories spectra with one astronomical IR spectrum (Young Stellar Object NGC 7538 IRS9 (Allamandola et al. 1992); Planetary Nebulae IRAS 05341+0852 (Joblin et al. 1996)); Murchison meteorite extract (de Vries et al. 1993). }\label{fig:compICES}
\end{figure}

Figure~\ref{fig:comp3p4micron} presents a comparison between the 3.4 $\mu$m band observed in four different astronomical objects: Young Stellar Object (YSO) NGC 7538 IRS9 (Allamandola et al. 1992); Planetary Nebulae (PN) IRAS 05341+0852 (Joblin et al. 1996); Proto Planetary Nebulae (PPN) CRL 618 (Chiar et al. 1998); and Diffuse Interstellar Medium (DISM) at galactic center through IRS 6E (Pendleton et al. 1994), and in the Murchison meteorite extract (de Vries et al. 1993) with the FTIR spectra of the studied H$_2$O:NH$_3$:c-C$_6$H$_{12}$ ice. Each set in Fig~\ref{fig:comp3p4micron}a shows the 3.4 $\mu$m band of bombarded H$_2$O:NH$_3$:c-C$_6$H$_{12}$ ice at three different ion fluences (0, 1.3$\times 10^{13}$, and 3$\times 10^{13}$ ions cm$^{-2}$). Each set of Fig.~\ref{fig:comp3p4micron}b illustrates the evolution of 3.4 $\mu$m band of bombarded H$_2$O:NH$_3$:c-C$_6$H$_{12}$ ice after 3$\times 10^{13}$ ions cm$^{-2}$ and at some temperatures during the heating from 13 K to room temperature. Temperature values are indicated in the figure.

The 3.46 $\mu$m ($\sim$2880 cm$^{-1}$) sharp feature observed in the YSO NGC 7538 IRS9 (squares in the upper set of Fig.~\ref{fig:comp3p4micron}a-b) indicates that cyclohexane (and possibly other aliphatic hydrocarbons) was not present initially in the ices of this protostellar object. As discussed by Allamandola et al. (1992), the peak at 3.54 $\mu$m ($\sim$2850 cm$^{-1}$) is attributed to methanol and this species is expected to be in the chemical inventory of protostellar grains.

The comparison between the 3.4 $\mu$m band in the PN IRAS 05341+0852 (the IR spectrum was inverted to appears like an absorption-type spectrum in Fig.~\ref{fig:comp3p4micron}) and the H$_2$O:NH$_3$:c-C$_6$H$_{12}$ ice is remarkably good. The -CH$_3$ stretching mode of aliphatic hydrocarbons at 3.37 and 3.48 $\mu$m is a good indicator of the amount of energy that was delivered to the grains. The same is also true for the =CH$_2$ (sp$^2$) vibration mode at 3.33 $\mu$m ($\sim$3010 cm$^{-1}$), which becomes very strong only at higher Fluences (e.g. 3$\times 10^{13}$ ions cm$^{-2}$). As illustrated in Fig.~\ref{fig:comp3p4micron}b, this band is very sensitive to the temperature of the ice, suggesting that the carrier must be a volatile alkene. This band may also be used as a thermometer for bombarded interstellar ices. In addition, since this band is not observed in Murchison meteorite extraction, this suggests that possible volatile alkene species that were present at one time are now completely lost. Looking at the temperature effect on the 3.33 $\mu$m  band (4th graph set in Fig.~\ref{fig:comp3p4micron}b), the interstellar grains in the DISM through IRS 6E should have experienced temperatures higher than 50 K.

\begin{figure*}
 \centering
\resizebox{\hsize}{!}{\includegraphics{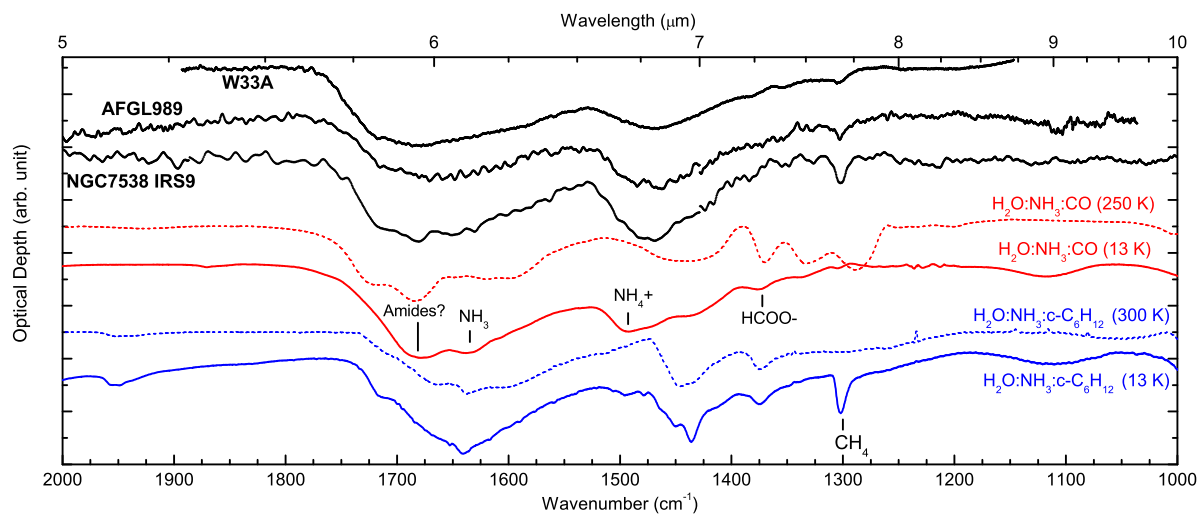}}
\caption{Comparison between IR spectra from 2000 to 1000 cm$^{-1}$ (6-10 $\mu$m) some of interstellar and laboratory ices. The top three curves are infrared spectra of young stellar sources obtained
by the Infrared Space Observatory (ISO). Lower traces indicate different
laboratory spectra of two ammonia-containing ices irradiated by heavy-ions
at 13 K (H$_2$O:NH$_3$:CO (1:0.6:0.4) bombarded with 46 MeV $^{46}$Ni$^{13+}$ (Pilling et al. 2010a) and H$_2$O:NH$_3$:c-C$_6$H$_{12}$ (1:0.3:0.7) bombarded with 632 MeV $^{58}$Ni$^{24+}$ (this work)).}
\label{fig:Astromomical}
\end{figure*}

In general, the 3.4 $\mu$m band in PPN CRL 618, DISM IRS 6E, and Murchison extract is very similar with each other, and by excepting the sharp absorption peak at 3.37 $\mu$m ($\sim$ 2960 cm$^{-1}$), they are also very similar to the IR spectra of bombarded H$_2$O:NH$_3$:c-C$_6$H$_{12}$ ice at a fluence of 3$\times 10^{13}$ ions cm$^{-2}$. The enhancement of the 3.37 $\mu$m band as a function of fluence suggests that these interstellar grains (and the grains in the Murchison meteorite) probably were exposed to a heavy cosmic ray fluence higher than 3$\times 10^{13}$ ions cm$^{-2}$ or to some kind of ionizing agents which delivered an energy dose higher than $\sim$ 2 $\times$ 10$^{8}$ J kg$^{-1}$ $\approx 10^{15}$ eV ng$^{-1}$\footnote{Considering a typical density of 1 g cm$^{3}$ and a size of 0.1 $\mu$m, the typical mass of interstellar grains is about 1 ng.} (10$^{13}$ ions cm$^{-2}$ $\sim$ 6$\times$ 10$^{7}$ J kg$^{-1}$; Pilling et al. 2011).

Figure~\ref{fig:compICES} presents a comparison of the 3.4 $\mu$m band observed after heavy-ion bombardment of two simulated interstellar ices at 13 K and after warming (H$_2$O:NH$_3$:CO (1:0.6:0.4) bombarded with 46 MeV $^{46}$Ni$^{13+}$ (Pilling et al. 2010a) and H$_2$O:NH$_3$:c-C$_6$H$_{12}$ (1:0.3:0.7) bombarded with 632 MeV $^{58}$Ni$^{24+}$ (this work). Both experiments were performed employing the same instrumentation and methodology. The ion fluences and ices temperatures are indicated in the figure. In each set of this figure we also present a comparison of the laboratories spectra with an astronomical IR spectrum (Young Stellar Object NGC 7538 IRS9 (Allamandola et al. 1992); Planetary Nebulae IRAS 05341+0852 (Joblin et al. 1996)); Murchison meteorite extract (de Vries et al. 1993). The 3.46 $\mu$m feature observed in the young stellar object NGC 7538 IRS9 is in a good agreement with the band detected in the radiolysis product from H$_2$O:NH$_3$:CO ice at 13 K. However, small features at 3.7 $\mu$m and 3.43 $\mu$m seem to match the infrared features of the 300 K residue for bombarded H$_2$O:NH$_3$:c-C$_6$H$_{12}$ ices. Furthermore, an additional abundance of methanol is needed to get a good fitting of the observation at 3.53 $\mu$m. This scenario suggests that the initial abundance of this protostelar disk may contain species such as NH$_3$, CO, and simple aliphatic hydrocarbon including cyclohexane and methanol as pointed out by Allamandola et al (1992). The data also indicates that warmer ices (closer to the star) may be rich in  simple aliphatic hydrocarbon and colder ices rich in CO.

For the planetary nebulae IRAS 05341+0852 (IR spectrum was inverted to appears like an absorption-type spectrum in Fig.~\ref{fig:compICES}), the best match occurred with the residue ($\sim$ 130 K) obtained after the ion bombardment of H$_2$O:NH$_3$:c-C$_6$H$_{12}$ ice. This puts a constrain on the temperature of dust grains in the region and indicates that a negligible amount of CO was in the grains initially. The organic residues for both bombarded ices (H$_2$O:NH$_3$:CO and H$_2$O:NH$_3$:c-C$_6$H$_{12}$) at temperatures hotter than 250 K have some similarities in this wavelength region with peak around 3.53 $\mu$m and 3.42$\mu$m associated with aliphatic hydrocarbons. However, the IR spectrum of Murchison extraction around 3.4$\mu$m can not be fitted by any spectrum taken from  these two simple interstellar ice analogs.

\subsection{5-10 $\mu$m range}

Figure~\ref{fig:Astromomical} presents a comparison between IR spectra from 5 to 10 $\mu$m (2000 to 1000 cm$^{-1}$) of interstellar and bombarded laboratory ices. The top three curves are infrared spectra of YSOs (W33, AFGL989, NGC7538 IRS9) obtained by the Infrared Space Observatory (ISO). The four bottom curves in this figure indicate different laboratory spectra of two ammonia-containing ices irradiated by heavy and energetic ions: H$_2$O:NH$_3$:CO (1:0.6:0.4) at fluence of 2$\times 10^{13}$ ion cm$^{2}$ (Pilling et al. 2010a), and H$_2$O:NH$_3$:c-C$_6$H$_{12}$ (1:0.3:0.7) at fluence of 3$\times 10^{13}$ ion cm$^{2}$ (this work). For comparison, the organic residues (after heating) of these two simulated ices after bombardment are also shown.

The 7.7 $\mu$ m (1300 cm$^{-1}$) feature in YSOs, which is attributed to CH$_4$ (-CH deformation mode; Boogert et al. 1997), is best represented by the products coming from the radiolysis of ices containing alkanes initially, rather than by the ices where the primary source of carbon comes from CO. This suggests that saturated hydrocarbons, such as cyclohexane, may be a part of the initial chemical inventory of interstellar ices. However, the presence of broad features at 1450 cm$^{-1}$ and 1680 cm$^{-1}$ in the astronomical observations, similar to the features observed in the IR spectra of the irradiated H$_2$O:NH$_3$:CO ice (Pilling et al. 2010a) requires the presence of CO (or/and CO$_2$) as one of the parental species in the interstellar grain ice mantles. Based on these results and with the fact that the 3.4 $\mu$m band in these YSOs requires the presence of methanol (e.g. Allamandola et al. 1992), we propose that the initial composition of grains in the disc of these young stellar sources should contain a mixture of H$_2$O, NH$_3$, CO or (CO$_2$), alkanes, and CH$_3$OH. A future work will focus on the initial abundance of these simple compound in young stellar environments.

\section{Conclusion} 

We performed an experimental study of the interaction of medium-mass ions (219 MeV $^{16}$O$^{7+}$) and heavy-ions ions (632 MeV $^{58}$Ni$^{13+}$) with cyclohexane-containing ices at 13 K (pure-c-C$_6$H$_{12}$ and  H$_2$O:NH$_3$:c-C$_6$H$_{12}$ (1:0.3:0.7)) to simulate the physico-chemical changes (e.g. molecular unsaturation) induced by cosmic rays in saturated hydrocarbon-rich ice analogs inside dense regions of the interstellar medium such as dense molecular clouds or protoplanetary disks, as well on the surfaces of solar system ices. Our main results and conclusions are the following.

\begin{enumerate}

 \item For the experiment with medium-mass cosmic ray analogs (219 MeV O$^{7+}$) on pure c-C$_6$H$_{12}$ ice at 13 K, the determined  dissociation cross section is 1 $\times 10^{-14}$ cm$^{2}$, and the sputtering yield is 10$^3$ c-C$_6$H$_{12}$ molecules per ion impact. For the experiment in which mixed H$_2$O:NH$_3$:c-C$_6$H$_{12}$ (1:0.3:0.7) ice at 13 K was bombarded by heavy-ion cosmic ray analogs (632 MeV Ni$^{24+}$), the obtained dissociation cross sections of H$_2$O, NH$_3$ and c-C$_6$H$_{12}$ are $\sim$3 $\times 10^{-13}$, $\sim$3 $\times 10^{-13}$, and $\sim$2 $\times 10^{-13}$ cm$^{2}$, respectively. The results indicate that the dissociation cross section of cyclohexane is about one order of magnitude higher when swift heavy cosmic ray analogs were employed in comparison with the experiment employing medium-mass cosmic rays.

 \item The half-life of cyclohexane ($t_{1/2}$) in ISM as a result of cosmic ray bombardment for pure c-C$_6$H$_{12}$ ice is found to be about $5 \times 10^6$ years (considering only medium-mass cosmic rays); for mixed ice it is about $1 \times 10^6$ years (considering only heavy-ion cosmic rays).

 \item For the mixed ice (H$_2$O:NH$_3$:c-C$_6$H$_{12}$), after 20 $\times 10^6$ years in ISM, almost 20\% of the initial cyclohexane was converted into CO by heavy cosmic rays, 3\% was transformed into OCN$^-$, and 1\% into CO$_2$. Such result suggests that highly hydrogenated hydrocarbons in water-rich grain mantles inside interstellar clouds (or other astrophysical ices exposed to medium-mass and heavy cosmic rays) can be mostly converted into CO.

 \item The results of the radiolysis of simple alkanes such as cyclohexane, in astrophysical ice analogs are consistent with the production of unsaturated molecules containing double bond(s), triple bond(s) (e.g. 2334 cm$^{-1}$), and/or ring(s) (e.g. 3085 cm$^{-1}$). Extrapolating to an astrophysical scenario, the maximum production of alkenes is obtained after about 3-5 $\times 10^6$ years independent of the projectile type and the presence of polar species in the ice (e.g. water and ammonia). The formation cross section of alkenes, considering the C-H out-of-plane bending mode at 719 cm$^{-1}$ ($A \sim 8\times 10^{-18}$ cm molec$^{-1}$), in the pure c-C$_{6}$H$_{12}$ ice and mixed H$_2$O:NH$_3$:c-C$_6$H$_{12}$ ice were about $1\times 10^{-15}$ and $1\times 10^{-14}$ cm$^{2}$. The maximum value was roughly the same for both experiments, $\sim$ 10$^{-2}$ alkenes per cyclohexane molecule irradiated by cosmic ray analog. Another consequence of this induced unsaturation is the production of molecular hydrogen in the ice, which can be released to the gas phase (desorbed) depending on its energy or ice temperature.

\item The 3.4 $\mu$m band of PN IRAS 05341+0852 has large similarity with the band observed for the organic residue ($\sim$ 130 K) obtained after the ion bombardment of H$_2$O:NH$_3$:c-C$_6$H$_{12}$ ice. This suggests two different issues: first, it puts a constrain in the temperature of dust grains in the region, and secondly, it also indicates that a negligible amount of CO was present in the grains initially.

\item Considering the sharp 3.37 $\mu$m absorption peak (C-H asymmetric stretching mode in -CH$_3$) we suggest that grains in PPN CRL 618, DISM GC IRS 6E, and Murchison carbonaceous meteorite were probably exposed to heavy cosmic rays with fluence higher than 3$\times 10^{13}$ ions cm$^{-2}$ or to some kind of ionizing agents that delivered an energy dose higher than $\sim$ 2 $\times$ 10$^{8}$ J kg$^{-1}$ $\approx 10^{15}$ eV ng$^{-1}$.

\item A comparison between IR spectra from the laboratory and YSOs suggests that the initial composition of grains in young stellar disks may include a mixture of H$_2$O, NH$_3$, CO (or CO$_2$), alkanes, and CH$_3$OH.

\end{enumerate}

Finally, the results indicated that cosmic ray bombardment of saturated alkanes can be an alternative scenario for the production of unsaturated hydrocarbons and possibly aromatic rings (via dehydrogenation processes) in interstellar and protostellar ices. Moreover, a comparison between the laboratory spectra and infrared observations of protostellar ices indicates that saturated hydrocarbons such as cyclohexane may be a part of the initial chemical inventory of interstellar ices.

%
\section*{Acknowledgments}
The authors acknowledge the agencies COFECUB (France), CAPES, CNPq, and FAPERJ (Brazil) for financial support. We also thank Ms. Alene Alder Rangel for the English revision of this manuscript.

%

\appendix
\section{New IR features after Radiolysis}

Table~\ref{tab:newspec} lists the new IR features formed after ion bombardment of the cyclohexane containing ices at 13 K.

\begin{table*}
\begin{center}
\caption{Possible assignment of infrared absorption features produced by the radiolysis of the cyclohexane containing ices at 13 K.} \label{tab:newspec}
\setlength{\tabcolsep}{6pt}
\begin{tabular}{ c c l l }
\hline
\multicolumn{2}{c}{Wavenumber (cm$^{-1}$)}    & Attribution and comments                     & References                   \\
\cline{1-2}
c-C$_6$H$_{12}$ & H$_2$O:NH$_3$:c-C$_6$H$_{12}$ &                                           &                      \\
\hline
-            &	3091                 & $=$C-H (sp$^{2}$) str. benzene; cyclohexene; alkenes      & [1,6]           \\
3021	     &  3021$^a$             & $=$C-H (sp$^{2}$) str. benzene; aromatic ?                & [1,6]          \\
-	         &  3008	             & $=$C-H (sp$^{2}$) str. alkenes                            & [1]          \\
-	         &  2971                 & ?                                                    &           \\
2341         &	2337	             & CO$_2$                                               & [2]          \\
2334         &  -                    & -C$\equiv$C- sym. str of non sym. alkynes            & [1]       \\
-	         &  2258                 & CH$_2$-C$\equiv$N? (C$\equiv$N str.)                 & [7]          \\
-	         &  2225	             & -C$\equiv$C-H (C$\equiv$C str. of internal alkynes at 2220 cm$^{-1}$)?  & [1]           \\
-	         &  2166                 & OCN-                                                 & [3]          \\
2133	     & 2137                  & CO; -C$\equiv$C-H (C$\equiv$C str. of terminal alkynes at 2120 cm$^{-1}$)? & [1]          \\
-	         & 2081                  & HCN                                                  &           \\	
- 	         & 1951                  & ?                                                   &           \\	
-	         & 1713                  & ?                                                   &           \\	
1688	     & -                     & broad. C$=$C str. aromatic?, ring?                  & [1]        \\
1642         & 1641	                 & broad. C$=$C str. aromatic?, ring?                  & [1]        \\
1584	     & 1584	                 & broad. NH$_3$CH$_2$COO$^-$?                         & [4]        \\
-            & 1436	                 & NH$_3$CH$_2$COO$^-$?                                & [4]        \\
1392	     & -                     & t-butyl; benzene?                                   & [1]       \\
1376	     & 1375                  &	-CH$_3$ bending sym. t-butyl or dimethyl           & [1]       \\
1300	     & 1302	                 & CH$_4$ (C-H $\nu_4$ deformation mode)               & [5]       \\
1137	     & 1136$^a$              & ?                                                   &           \\
993  	     & 994	                 & $_H^R$C$=$C$^H_H$  (C-H bending out plane); cyclohexene; aromatic? & [1]          \\
967	         & -	                 & ?                                                   &           \\
-	         & 954	                 & ?                                                   &           \\
948	         & -	                 & ?                                                   &           \\
916	         & 916	                 & $_H^R$C$=$C$^H_H$  (C-H bending out plane); terminal vinyl at 910 cm$^{-1}$? & [1]  \\
877	         & 877$^a$               & ?                                                   &           \\
-            & 845	                 & $_R^R$C$=$C$^R_H$  (C-H bending out plane)          & [1]       \\
818	         & -                     & ? weak                                              &           \\
810	         & -                     & ? weak                                              &           \\
751	         & 754	                 & ? broad and strong; aromatic C-H bending out plane; & [1]          \\
             &                       & CH$_2$ rocking aliphatic alkanes (present in molecules with less then 4 carbon  &    \\
             &                       & and very strong in molecules with less then 3 carbon)                    &              \\
720	         & 719$^a$               & $_H^R$C$=$C$^R_H$  bending out of plane aliphatic alkenes; cyclohexene   & [1]          \\
680	         & 683	                 & benzene?  (C-H bending in plane)                                          & [1,6]        \\
661		     & -                     & CO$_2$  bending?                                          & [2]         \\
643	         & 644	                 & $_H^R$C$=$C$^R_H$  (C-H bending out of plane); cyclohexene & [1]          \\
\hline

\multicolumn{4}{l}{[1] Pavia et al. 2009; [2] Gerakines et al. 1995; [3] Demyk et al. (1998);  [4] Holtom et al. 2005;}\\
\multicolumn{4}{l}{[5] Boogert et al. 1997; [6] Ruiterkamp et al. 2005; [7] Larkin 2011.}\\
\multicolumn{4}{l}{ $^a$absent for fluences $ \geq  3\times 10^{-13}$ ions cm$^{-2}$.}\\
\end{tabular}
\end{center}
\end{table*}

\bsp 
\label{lastpage}

\begin{thebibliography}{99}

\bibitem[\protect\citeauthoryear{Allamandola et al.}{1992}]{Allamandola92} 
Allamandola L.J., Sandford S.A., Tielens A.G.G.M., Herbst T.M., 1992, ApJ, 399, 134.

\bibitem[\protect\citeauthoryear{Apponi et al.}{2000}]{appomni_etal2000} %
Apponi A.J., McCarthy M.C., Gottlieb C.A.,  Thaddeus P., 2000, ApJ, 530, 357.

\bibitem[\protect\citeauthoryear{Bar-Num et al.}{1980}]{bar_etal80} 
Bar-Num A., Litman M. \& Rappaport M.L., 1980, A\&A, 85, 197.


\bibitem[\protect\citeauthoryear{Boogert et al.}{1997}]{Boogert_etal1997} %
Boogert A.C.A., Schutte W.A., Helmich F.P., Tielens A.G.G.M. \& Wooden D.H., 1997, A\&A, 317, 929.

\bibitem[\protect\citeauthoryear{Brown et al. }{1984}]{brow1984}
Brown W.L., Augustyniak W.M.,  Marcantonio K.J., Boring J.W., Johnson R.E.,  Reimann C.T., 1984, Nucl. Instrum. Methods Phys. Res., Sect. B, IV, 307.

\bibitem[\protect\citeauthoryear{Cernicharo et al. }{2001a}]{cernicharo2001a}
Cernicharo J., Heras A.M., Tielens A.G.G.M., Pardo J.R., Herpin F., Guélin M., Waters L.B.F.M., 2001a, ApJ, 546, L123.

\bibitem[\protect\citeauthoryear{Cernicharo et al. }{2001a}]{cernicharo2001a}
Cernicharo J., Heras A. M., Pardo J.R., Tielens A.G.G.M., Guélin D. E., Neri R., Waters L.B.F.M., 2001b, ApJ, 546, L127.

\bibitem[\protect\citeauthoryear{Cherchneff et al. }{1992}]{cherc1992}
Cherchneff I., Barker J.R., Tielens A.G.G.M., 1992, ApJ, 401, 269.

\bibitem[\protect\citeauthoryear{Chiar et al.}{1998}]{Chiar_etal1998} 
Chiar J.E., Pendleton, Y. J., Geballe T.R., \& Tielens A.G.G.M., 1998, ApJ, 507, 281.

\bibitem[\protect\citeauthoryear{Cordiner and Millar}{2009}]{cord2009} %
Cordiner M. A. \& Millar T.J., 2009, ApJ, 697, 68.

\bibitem[\protect\citeauthoryear{de Barros et al.}{2011}]{deBarros2011} 
de Barros A.L.F., Bordalo V., Seperuelo Duarte E., da Silveira E.F.,
 Domaracka A., Rothard H. \& Boduch P., 2011, A\&A, 531, A160.

\bibitem[\protect\citeauthoryear{Demyk et al. }{1998}]{demyk_etal1998}
Demyk K., Dartois E., D'Hendecourt L., Jourdain de Muizon M., Heras A.M., Breitfellner, 1998, A\&A, 339, 553.

\bibitem[\protect\citeauthoryear{de Vries et al.}{1993}]{devries_etal1993} 
de Vries M.S., Reihs K., Wendt H.R., Golden W.G., Hunziker H., Flemming R., Peterson E. \& Chang S., 1993, Geochim. Cosmochim. Acta, 57, 933.

\bibitem[\protect\citeauthoryear{d´Hendecourt and Allamandola}{1986}]{hend_allam1986} %
d´Hendecourt L. B. \& Allamandola L. J., 1986, Astron. Astrophys. Suppl. Ser. 64, 453.

\bibitem[\protect\citeauthoryear{d´Hendecourt et al.}{1986}]{hend_etal1986}
d'Hendecourt L.B., Allamandola L.J., Grim R.J.A., Greenberg J.M. 1986, A\&A, 158, 119.

\bibitem[\protect\citeauthoryear{Duley et al. }{2005}]{duley2005}
Duley W.W., Grishko V.I., Kenel J., Lee-Dadswell G., Scott A., 2005, ApJ, 626, 933.

\bibitem[\protect\citeauthoryear{Duley et al. }{2006}]{duley2006}
Duley W.W., Grishko V.I., Pinho G., 2006, ApJ, 642, 966.

\bibitem[\protect\citeauthoryear{Ehrenfreund \& Charnley}{2000}]{Ehren2000}
Ehrenfreund P. \& Charnley S.B., 2000, ARA\&A, 38, 427.

\bibitem[\protect\citeauthoryear{Frenklach and Feigelson}{1989}]{FrenklachFeigelson1989} 
Frenklach M. \& Feigelson E.D., 1989, ApJ, 341, 372.

\bibitem[\protect\citeauthoryear{Fukuzawa et al.}{1998}]{fukuzawa98} 
Fukuzawa K., Osamura Y., Schaefer III H. F., 1998, ApJ, 505, 278.

\bibitem[\protect\citeauthoryear{Gerakines et al. }{1995}]{gerakines_etal1995} %
Gerakines P.A., Scutte W.A., Greenberg J.M., van Dishoeck E. F., 1995, A\&A, 296, 810.

\bibitem[\protect\citeauthoryear{Gerakines et al. }{2005}]{Gerakines2005} 
Gerakines P.A., Bray J.J., Davis A. \& Richey C.R., 2005, ApJ, 620, 1140.

\bibitem[\protect\citeauthoryear{Herbst and van Dishoeck}{2009}]{HerbstDishoeck2009} %
Herbst E., van Dishoeck E.F., 2009, ARA\&A, 47, 427.

\bibitem[\protect\citeauthoryear{Herbst}{1995}]{Herbst95} %
Herbst E., 1995, Annu. Rev. Phys. Chem., 46, 27.

\bibitem[\protect\citeauthoryear{Holtom et al.}{2005}]{Holtom05} 
Holtom P.D., Bennett C.J., Osamura Y., Mason N.J. \& Kaiser R.I.,
2005, ApJ, 626, 940.


\bibitem[\protect\citeauthoryear{Iglesias-Groth et al. }{2010}]{igle2010}
Iglesias-Groth S., Manchado A., Rebolo R., González Hernández J.I., García-Hernáandez D.A. \& Lambert D.L., 2010, MNRAS, 407, 2157.

\bibitem[\protect\citeauthoryear{Iglesias-Groth et al. }{2008}]{igle2008}
Iglesias-Groth S., Manchado A., González Hernández D.A., García-Hernández J.I. \& Lambert D.L., 2008, ApJ, 685, L55.

\bibitem[\protect\citeauthoryear{Jiang et al.}{1975}]{jiang_etal75} 
Jiang G.J., Pearson W.B., Brown K.G., 1975, J. Chem. Phys., 65, 1201.

\bibitem[\protect\citeauthoryear{Joblin et al.}{1996}]{joblin_etal96}
Joblin C., Tielens A.G.G.M., Allamandola L.J., \& Geballe T.R., 1996,
ApJ, 458, 610.

\bibitem[\protect\citeauthoryear{Kerkhof et al. }{1999}]{Kerkhof1999} 
Kerkhof O., Schutte W.A. \& Ehrenfreund P., 1999, A\&A, 346, 990.

\bibitem[\protect\citeauthoryear{Larkin}{2011}]{Larkin2011} 
Larkin P.J. (2011) in Infrared and raman spectroscopy: principles and spectral interpretation. Elsevier Inc. USA.

\bibitem[\protect\citeauthoryear{Loeffler et al.}{2005}]{Loeffler_etal2005} 
Loeffler M.J., Barata G. A., Palumbo M.E., Strazulla G., Baragiola R.A., 2005, A\&A, 435, 587.

\bibitem[\protect\citeauthoryear{Nastasi et al. }{1996}]{nastasi_etal1996} 
Nastasi M., Mayer J. \& Hirvonen J.K., 1996, in ``Ion-Solid Interactions: Fundamentals and Applications'', Cambridge Solid State Science Series, Cambridge University Press

\bibitem[\protect\citeauthoryear{Marcelino et al. }{2007}]{marcelino_etal07} 
Marcelino N., Cerhicharo J., Agúndez M., Roueff E., Gerin M., Martín-Pintado J., Mauersberger R. \& Thum C., 2007, ApJ, 665, L127.

\bibitem[\protect\citeauthoryear{Olah and Molnár. }{2003}]{OlahBOOK} 
Olah G.A. \& Molnár A., 2003, in ``Hydrocarbon Chemistry'', 2nd Edition, Wiley-Interscience, USA.

\bibitem[\protect\citeauthoryear{Pavia et al. }{2009}]{paviaBOOK} 
Pavia L. D., Lampman G. M., Kriz G. S., Vyvyan J. A., 2009, Introduction to Spectroscopy, 4th Edition, Cengage Learning, USA.

\bibitem[\protect\citeauthoryear{Pendleton et al.}{1994}]{pendleton94} 
Pendleton Y.J., Sandford S.A., Allamandola L.J., Tielens A.G.G.M. \& Sellgren K., 1994, ApJ, 437, 683.

\bibitem[\protect\citeauthoryear{Pendleton and Allamandola}{2002}]{Plend_allam2002}
Pendleton Y.J. \& Allamandola L.J., 2002, APJSS, 138, 75.

\bibitem[\protect\citeauthoryear{Pilling et al.}{2010a}]{pilling_etal2010a} 
Pilling S., Seperuelo Duarte E., da Silveira E.F., Balanzat E., Rothard H., Domaracka A., Boduch P., 2010a, A\&A, 509, A87.

\bibitem[\protect\citeauthoryear{Pilling et al.}{2010b}]{pilling_etal2010b} 
Pilling S., Seperuelo Duarte E.,  Domaracka A, Rothard H., Boduch P., da Silveira E.F., 2010b, A\&A 523, A77.

\bibitem[\protect\citeauthoryear{Pilling et al.}{2011}]{pilling2011} 
Pilling S., Seperuelo Duarte E., Domaracka A., Rothard H., Boduch P., da Silveira E. F., 2011, Phys. Chem. Chem. Phys., 13, 15755.

\bibitem[\protect\citeauthoryear{Ruiterkamp et al.}{2005}]{Ruiterkamp2005}  
Ruiterkamp R., Peeters Z., Moore M.H., Hudson R.L. \& Ehrenfreund P., 2005, A\&A, 440, 391.

\bibitem[\protect\citeauthoryear{Seperuelo Duarte et al. }{2009}]{seperuelo_etal2009} 
Seperuelo Duarte E., Boduch P., Rothard H., Been T., Dartois E., Farenzena L.S. \& da Silveira E. F. 2009, A\&A, 502, 599.

\bibitem[\protect\citeauthoryear{Seperuelo Duarte et al. }{2010}]{seperuelo_etal2010} 
Seperuelo Duarte E., Domaracka A., Boduch P., Rothard H., Dartois E. \& da Silveira E. F. 2010, A\&A, 512, A71.

\bibitem[\protect\citeauthoryear{Sirjean et al.}{2006}]{Sirjean_etal2006} %
Sirjean B., Glaude P. -A., Ruiz-Lopez M. F., Fournet R., 2006, J. Physical Chemistry A, 110, 12693.

\bibitem[\protect\citeauthoryear{Smith}{1992}]{Smith1992} %
Smith D., 1992, Chem. Rev., 92, 1473.

\bibitem[\protect\citeauthoryear{Taylor and Duley}{1997}]{tayDuley97} 
Taylor S.D. \& Duley W.W., 1997, MNRAS, 286, 344.

\bibitem[\protect\citeauthoryear{Thaddeus and McCarthy}{2011}]{ThadMacarty2011} %
Thaddeus P. \&  McCarthy M.C., 2011, Spectrochim. Acta A, 57, 757.

\bibitem[\protect\citeauthoryear{Winnewisser and Herbst}{1987}]{Winnewisser87} %
Winnewisser G. \& Herbst E. 1987, Topics Current Chem., 139, 119.

\bibitem[\protect\citeauthoryear{Ziegler et al.}{2011}]{zieg2011}
Ziegler J.F., Ziegler M.D., Biersack J.P., 2011, Stopping and Range of Ions in Matter - SRIM2011 (available at www.srim.org).


\end{thebibliography}
\end{document}